\documentclass[a4paper,12pt,oneside]{article}
\usepackage[utf8]{inputenc}
\usepackage{graphicx}
\usepackage[T1]{fontenc}
\usepackage{amsmath}
\usepackage{amssymb}
\usepackage{float}
\usepackage{hyperref}
\usepackage{cleveref}
\usepackage{lmodern}
\usepackage{empheq}
\usepackage{braket}
\usepackage{xcolor}
\usepackage{cancel}
\usepackage[pagewise]{lineno}
\usepackage{csquotes}
\usepackage{scalerel}
\usepackage{appendix}
\usepackage{epigraph}
\usepackage{afterpage}
\usepackage[mathscr]{eucal}
\usepackage[shortlabels]{enumitem}
\usepackage{import}
\usepackage{tikz}
\usetikzlibrary{shapes.geometric}
\usepackage{booktabs}
\usepackage{mathtools}
\usepackage{titlesec}
\usepackage{caption}
\usepackage{subcaption}
\usepackage{textgreek}
\usepackage{blindtext}
\usepackage{titlesec}
\usepackage{indentfirst}

\titleformat{\section}
{\normalfont\fontsize{12}{12}\selectfont\bfseries} 
{\thesection}
{1em}
{}

\titleformat{\subsection}
{\normalfont\fontsize{12}{12}\selectfont\bfseries} 
{\thesubsection}
{1em}
{}

\newcommand{\orcid}[1]{\href{https://orcid.org/#1}{\includegraphics[width=8pt]{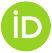}}}

\usepackage[left=3.0cm,right=2.5cm,top=3.0cm,bottom=2.5cm]{geometry}

\hypersetup{
	colorlinks   = true, %Colours links instead of ugly boxes
	urlcolor     = blue, %Colour for external hyperlinks
	linkcolor    = blue, %Colour of internal links
	citecolor   = blue %Colour of citations
}

\begin{document}
	
\numberwithin{equation}{section}

	\begin{center}
		\vspace{5mm}
		\large
		\textbf{Emergence of Dark Phases in Scalar Particles within the Schwarzschild-Kiselev-Letelier Spacetime}\\ %Só coloquei um título arbitrário, deve ser alterado :)
		\normalsize
		\vspace{5mm}
		B. V. Simão\footnote{vallin.bruna@posgrad.ufsc.br}\,\orcid{0000-0001-6087-5608}\,,
        M. L. Deglmann\textsuperscript{*}\footnote{m.l.deglmann@posgrad.ufsc.br}\,\orcid{0000-0002-9737-5997}\,,
		C. C. Barros Jr.\footnote{barros.celso@ufsc.br}\,\orcid{0000-0003-2662-1844}.
        \let\thefootnote\relax\footnotetext{*\ Corresponding author: m.l.deglmann@posgrad.ufsc.br.}
	\end{center}
	
	\begin{center}
		Departmento de Física, Universidade Federal de Santa Catarina (UFSC), Campus Universitário Trindade, Florianópolis, 88035-972, Santa Catarina, Brazil
	\end{center}

\vspace{5mm}

\begin{abstract}
    This work focuses on the emergence of dark phases (dark energy-induced phases) in the radial wave function of scalar particles. We achieve this by presenting novel solutions to the Klein-Gordon equation in a spherically symmetric spacetime, which encompasses a black hole, a quintessential fluid, and a cloud of strings. We determine the exact solution for the spacetime metric, analyze the admissible ranges for its physical parameters, and discuss the formation of the event horizon. 
    Subsequently, we detail the solution of the Klein-Gordon equation and explore three distinct cases of dark phases, corresponding to the quintessence state parameter $\alpha_{\scaleto{Q}{4pt}}$ taking the values $0$, $1/2$, and $1$. Notably, the case where $\alpha_{\scaleto{Q}{4pt}} = 1$ holds particular significance due to current observational constraints on dark energy.\\
    %\vspace{5pt}
    
    \noindent \textbf{Keywords:} Dark Energy, Dark Phases, Quintessence, Black Holes, Spin-$0$ Particles, Heun Equations.
\end{abstract}

\section{Introduction}

Over the past decades, multiple independent observations have strongly supported a dark energy-driven cosmology. For example, groundbreaking data from the luminosity distances of Type Ia supernovae \cite{Riess1998, Perlmutter1999} and evidence from the cosmic microwave background (CMB), baryon acoustic oscillations (BAO), and large-scale structure (LSS) corroborate the universe's accelerated expansion \cite{Amendola2010,Steinhardt2003}. However, the nature of dark energy remains an open question, with a collective effort underway to determine its equation of state \cite{Escamilla2023}.

In $\Lambda$CDM cosmology, the cosmological constant ($\Lambda$) acts as the dark component, yet it faces substantial challenges \cite{Amendola2010}, which can be addressed by alternative dark energy models. In fact, recent findings from the Dark Energy Survey (DES) \cite{DESCollaboration2025} and the Dark Energy Spectroscopic Instrument (DESI) \cite{DESI62024,DESI2025} suggest that dark energy might have a dynamical nature. These observations point to a dark energy state parameter ($\omega_{\scaleto{Q}{4pt}}$) with values in the vicinity of $\omega_{\scaleto{Q}{4pt}}\approx -1$, a regime that includes even the so-called phantom dark energy. This observational evidence favors time-evolving models, particularly the class of modified matter models for dark energy, such as the original quintessence model \cite{Caldwell1998} and its subsequent developments in scalar field quintessence \cite{Amendola2010,Steinhardt2003}. Alternative frameworks have also been proposed \cite{Farrah_2023}, suggesting that black holes contribute cosmologically as vacuum energy. This feature leads to the intriguing possibility that black holes provide a nearly constant cosmological energy density, and dark energy itself could be an effect arising from them.

In contrast to the works that followed \cite{Caldwell1998}, Kiselev \cite{Kiselev2003} proposed an alternative approach, also termed \enquote{quintessence}, which we understand as an anisotropic, time-independent fluid with positive energy density and negative pressures. This approach offers a simpler way to investigate dark energy. Furthermore, although it lacks a time-evolving state parameter, Kiselev's quintessence can describe particular cosmic epochs where temporal variations might be negligible. Thus, we can use it to study the physical outcomes of a specific cosmic epoch where the time variations are negligible (see, for instance, \cite{Wang2023,Rayimbaev2024}).

This particular dark energy model also offers a unique opportunity to explore how spacetime curvature, specifically its quintessence component, affects quantum systems. This is a subject of great interest because we want to know whether a physical observable can be proposed for this purpose. Since Parker's formulation \cite{Parker:1980hlc}, we have explored this general question (of how geometry of space-time affects quantum mechanics) in various scenarios, for example, by investigating scalar bosons and Dirac particles \cite{Ahmed2020, elizalde_1987, Pinho2023, Santos2018, Santos2017, Soares:2021uep, Vitoria:2018mun, chandra, Guvendi:2022uvz,Sedaghatnia:2019xqb}. However, the question remains far from conclusive.

Building upon this framework, our previous work \cite{Deglmann2025} analyzed the dynamics of scalar particles in an anti-de Sitter (AdS) spacetime that contains a black string (BS) \cite{Lemos1995}, an analogue to clouds of strings (CS) \cite{Letelier1979}, and Kiselev's quintessence fluid \cite{Kiselev2003}. In these investigations, we primarily aimed to determine quintessence's effect on quantum systems (specifically, spin-$0$ particles) within a cylindrically symmetric spacetime. We successfully demonstrated that the quintessence fluid induces a possible quantum observable, which we termed the \enquote{dark phase}. These dark phases allow us to distinguish between different values of the dark energy state parameter in the Klein-Gordon radial solution. As a consequence, our findings highlight the significant role that even static quintessential fluids can play at the quantum level. Consequently, we anticipate that further investigation of these dark phases will yield valuable insights into the fundamental nature of dark energy and its quantum implications.

In the present work, we investigate the emergence of dark phases for spin-$0$ particles within a spherically symmetric spacetime. Specifically, we analyze the Schwarzschild-Kiselev-Letelier spacetime, which includes a black hole, the contribution of the Kiselev quintessential fluid, and a cloud of strings. We solve the Klein-Gordon equation and determine the occurrence of dark phases for three distinct values of the auxiliary state parameter $\alpha_{\scaleto{Q}{4pt}}=-(3\omega_{\scaleto{Q}{4pt}}+1)/2$, namely $\alpha_{\scaleto{Q}{4pt}}=0,\,1/2,\,1$. Since recent observational findings point to an equation of state parameter close to $\omega_{\scaleto{Q}{4pt}}=-1$ (which is equivalent to $\alpha_{\scaleto{Q}{4pt}}=1$), our analysis of this limiting case becomes particularly relevant as it represents the regime closest to the physical reality. These results, together with our previous ones in cylindrical symmetry, will enable us to compare the roles of both quintessence and spacetime symmetry in inducing phase shifts, depending on the quintessence parameter $N_{\scaleto{Q}{4pt}}$.

For this purpose, we have organized the paper as follows: In Section 2, we introduce the exact metric solution for the Schwarzschild-Kiselev-Letelier spacetime, along with its corresponding energy density and pressure. Section 3 provides parameter estimates for the quintessential component, as well as for the cloud parameter and the Schwarzschild radius. In Section 4, we analyze event horizon formation. The subsequent sections of the paper focus on our main results. In Section 5, we describe the solutions to the Klein-Gordon equation, which we then use to derive the dark phases presented in Section 6. As these results critically depend on the properties of the confluent Heun equation, we provide a detailed discussion of its local solution via the Frobenius method in the Appendix. Finally, we offer our concluding remarks and comparisons in Section 7.

\section{Einstein equations} \label{Section2}

In this work, we aim to study the existence of dark phases and their relevance concerning the other components of the metric. To achieve this, we assume a system composed of a central mass, a cloud of strings, and a Kiselev quintessence fluid. Therefore, in this section, we will determine the metric and the components of the energy-momentum tensor that lead to the desired formulation.
  
Let us consider the following interval
\begin{equation}
	ds^{2} = f(r)\,c^{2}dt^{2} - \frac{d r^{2}}{f(r)} - r^{2} d\theta^{2} - r^{2} \sin^{2}\theta\,d\varphi^{2}\,,\label{metric}
           \end{equation}
associated with the ansatz of a spherically symmetric spacetime, with a \emph{mostly minus} metric signature, related to a Schwarzschild black hole. We take the general Einstein tensor to be
\begin{equation}
	G_{\mu\nu} = R_{\mu\nu} - \frac{1}{2}g_{\mu\nu}\,R\,,
\end{equation}
where $R_{\mu\nu}$ is the Ricci tensor and $R$ is the Ricci scalar. Then, for the metric tensor associated with \cref{metric}, the spherical components $G_{\mu}^{\ \nu}$ of the Einstein tensor are
\begin{align}
	G_{t}^{\ t} = G_{r}^{\ r}
	&= \frac{1}{r}\,\frac{d f (r)}{d r} + \frac{f(r)}{r^{2}} - \frac{1}{r^{2}}\,,\label{Einstein_t_ut}\\[5pt]
	G_{\theta}^{\ \theta} = G_{\varphi}^{\ \varphi}
	&= \frac{1}{2}\,\frac{d^{2}f (r)}{d r^{2}} + \frac{1}{r}\,\frac{d f(r)}{d r}\,.\label{Einstein_theta_utheta}
\end{align}
We consider the Einstein equations to be:
\begin{equation}
	G_{\mu}^{\ \nu} = \frac{8\pi G}{c^{4}}\,T_{\mu}^{\ \nu}\,.\label{general-Einstein-equations}
\end{equation}
It is also important to note that we have chosen to work in the SI units convention to facilitate the interpretation of the theory's physical parameters. 

Considering the components of the Einstein tensor given by \cref{Einstein_t_ut,Einstein_theta_utheta}, along with the Einstein equations \eqref{general-Einstein-equations}, we see that any energy-momentum tensor compatible with this ansatz must be diagonal and possess the same symmetry. Its non-trivial components must satisfy:
\begin{equation}
    \begin{aligned}	
	T_{t}^{\ t} &= T_{r}^{\ r}\,,\\
	T_{\theta}^{\ \theta} &= T_{\varphi}^{\ \varphi}\,.\\
    \end{aligned} \label{General-T-munu}
\end{equation}
In other words, the Einstein equations impose a specific symmetry requirement on the energy-momentum tensor.

In this work, we consider the energy-momentum tensor of quintessence as:
\begin{equation}
    \begin{aligned}
        T_{t}^{\ t} &=  T_{r}^{\ r} = \rho_{\scaleto{Q}{4pt}}\,,\\
        T_{\theta}^{\ \theta} &=  T_{\varphi}^{\ \varphi} = \alpha_{\scaleto{Q}{4pt}}\,\rho_{\scaleto{Q}{4pt}}\,.
    \end{aligned} \label{Quintessence-Energy-Momentum-Tensor}
\end{equation}
This tensor is compatible with \cref{General-T-munu}, and here $\alpha_{\scaleto{Q}{4pt}}$ is defined by:
\begin{equation}
    \alpha_{\scaleto{Q}{4pt}} = -\frac{1}{2}\left(3\, \omega_{\scaleto{Q}{4pt}} + 1\right)\,, \label{eq:Alpha_Q_Definition}
\end{equation}
taking values in the interval $(0,1)$ since the standard state parameter $\omega_{\scaleto{Q}{4pt}}$ satisfies $\omega_{\scaleto{Q}{4pt}} \in (-1, -1/3)$. 

Additionally, we can incorporate a cloud of strings \cite{Letelier1979} into this system, whose energy-momentum tensor is given by:
\begin{equation}
    \begin{aligned}
        T_{t}^{\ t} &=  T_{r}^{\ r} = \frac{a}{r^{2}}\,,\\
        T_{\theta}^{\ \theta} &=  T_{\varphi}^{\ \varphi} = 0\,,
    \end{aligned} \label{energy-momentum-string-clouds}
\end{equation}
where $a$ is a real and positive constant ($a \in \mathbb{R}^{+}$).

With these two sources, the Einstein \eqref{general-Einstein-equations} equations become:
\begin{align}
    &\frac{1}{r}\,\frac{d f (r)}{d r} + \frac{f(r)}{r^{2}} - \frac{1}{r^{2}}
    = \frac{8\pi\,G}{c^{4}}\,\left(\rho_{\scaleto{Q}{4pt}} + \frac{a}{r^{2}}\right)\,\label{Einstein_Eq_1}\,,\\[5pt]
    &\frac{1}{2}\,\frac{d^{2}f (r)}{d r^{2}} + \frac{1}{r}\,\frac{d f(r)}{d r} 
    = \frac{8\pi\,G}{c^{4}}\, \alpha_{\scaleto{Q}{4pt}}\, \rho_{\scaleto{Q}{4pt}}\,.\label{Einstein_Eq_2}
\end{align}
To solve \cref{Einstein_Eq_1,Einstein_Eq_2}, we multiply \cref{Einstein_Eq_1} by $\alpha_{\scaleto{Q}{4pt}}$ and subtract it from \cref{Einstein_Eq_1}, to show that:
\begin{equation}
    r^{2}\,\frac{d^{2} f}{d r^{2}} + 2\left(1-\alpha_{\scaleto{Q}{4pt}}\right) r\,\frac{d f}{d r} - 2\alpha_{\scaleto{Q}{4pt}}\, f(r) = -2 \alpha_{\scaleto{Q}{4pt}} \left(1 + \frac{8\pi G}{c^{4}}\,a\right)\,.\label{Eq_f(r)}
\end{equation}
The associated homogeneous equation is in the form of a Cauchy-Euler equation, implying a solution of the form:
\begin{equation}
    f_{\scaleto{H}{4pt}} = -\frac{r_{s}}{r} + N_{\scaleto{Q}{4pt}}\, r^{2 \alpha_{\scaleto{Q}{4pt}}}\,,
\end{equation}
with $r_{s},\,N_{\scaleto{Q}{4pt}} \in \mathbb{R}^{+}$. In addition, the particular solution to \cref{Eq_f(r)} is simply $f_{\scaleto{P}{4pt}} = 1 + 8\pi G\,a/c^{4}$ so that the metric function $f(r)$ is determined by:
\begin{equation}
    f(r) = 1 + \overline{a} -\frac{r_{s}}{r} + N_{\scaleto{Q}{4pt}}\, r^{2 \alpha_{\scaleto{Q}{4pt}}}\,.
    \label{Metric_Function}
\end{equation}
Here, $\overline{a}$ is the dimensionless cloud parameter defined by:
\begin{equation}
    \overline{a} = \frac{8\pi G}{c^{4}}\,a\,.\label{Def_a_bar}
\end{equation}
The integration constants $r_{s}$ and $N_{\scaleto{Q}{4pt}}$ are the Schwarzschild radius and the quintessence parameter, respectively. To guarantee that $f(r)$ is dimensionless, we see that $r_{s}$ has dimension of length (i.e., meters in the SI units), while the dimension of $N_{\scaleto{Q}{4pt}}$ varies with $\alpha_{\scaleto{Q}{4pt}}$, satisfying:
\begin{equation}
	[N_{\scaleto{Q}{4pt}}] = [r^{-2\alpha_{\scaleto{Q}{3pt}}}] = \text{m}^{-2\alpha_{\scaleto{Q}{3pt}}}\,.\label{Dimension_N_Q}
\end{equation}
This means that when $\alpha_{\scaleto{Q}{4pt}}=0,\,1/2,\,1$ the quintessence parameter has dimensions of m$^{0}$, m$^{-1}$, and m$^{-2}$, respectively.

\subsection{Quintessence energy density and pressure}\label{energy-density-pressure}
	
Since we have determined $f(r)$, given by \cref{Eq_f(r)}, we use the first Einstein equation, \eqref{Einstein_Eq_1}, to obtain the quintessence energy density $\rho_{\scaleto{Q}{4pt}}(r)$, which yields:
\begin{equation}
\rho_{\scaleto{Q}{4pt}}(r) = \left(2\, \alpha_{\scaleto{Q}{4pt}} + 1\right)\frac{N_{\scaleto{Q}{4pt}}}{8\pi G/c^{4}}\, r^{2\, \alpha_{\scaleto{Q}{3pt}}-2}\,.\label{Energy_Density}
\end{equation}
Given that the energy density must be positive, we can see that the condition $N_{\scaleto{Q}{4pt}}\geq 0$ is compatible with the range of values $1<2\, \alpha_{\scaleto{Q}{4pt}} +1<3$. The results for the pressures $p_{\theta} = p_{\varphi}$ follow from \cref{Quintessence-Energy-Momentum-Tensor} and are given by
\begin{equation}
p_{\theta} = p_{\varphi} = -\frac{\alpha_{\scaleto{Q}{4pt}}\left(2\, \alpha_{\scaleto{Q}{4pt}} + 1\right)}{8\pi G/c^{4}}\,N_{\scaleto{Q}{4pt}}\,r^{2\, \alpha_{\scaleto{Q}{4pt}}-2}\,.\label{Pressures}
\end{equation}
As expected, $p_{\theta}$ and $p_{\varphi}$ are indeed negative. 

Our next task is to estimate the range of values for these physical parameters to gain a better understanding of our background spacetime.

\section{Estimating the physical parameters}\label{section-Estimates}

To estimate admissible values for the parameters occurring in the metric function $f(r)$ -- namely $\overline{a}$, $r_{\scaleto{s}{4pt}}$, and $N_{\scaleto{Q}{4pt}}$ -- it is useful to consider their dimensions in SI units and relate them with examples from stellar objects. For interested readers, similar approximations were detailed in our previous work \cite{Deglmann2025}.

We start by recalling that $f(r)$ is dimensionless. From this, it immediately follows that the Schwarzschild radius, $r_{s}$, has the dimension of length, i.e., $[r_{\scaleto{s}{4pt}}] = $ m, in SI units. Typical values for $r_{s}$ range from $10^{-3}$ m to $10^{15}$ m.

Furthermore, we emphasize that $\overline{a}$ is also dimensionless, with admissible values ranging from $10^{-6}$ to $10^{0}$, analogous to our findings in \cite{Deglmann2025}. The parameter $a$, appearing in \cref{energy-momentum-string-clouds}, has dimensions of $[a] = \text{kg m s}^{-2}$ in SI units. This is derived from its definition in \cref{Def_a_bar}, combined with the dimensions of the fundamental physical constants \cite{ParticleDataGroup2024, CODATA2022}.

Our next step is to estimate the values of $N_{\scaleto{Q}{4pt}}$, the quintessence parameter, in terms of the state parameter $\alpha_{\scaleto{Q}{4pt}}$. The strategy for this estimation is described in the following subsection.

\subsection{The quintessence parameter}

To estimate the range of values and the behavior for the quintessence parameter $N_{\scaleto{Q}{4pt}}$, we shall first approximate the amount of dark energy in our observable $\Lambda$CDM Universe. Let us consider the current radius of the observable Universe as $r_{\text{obs}}$ and the current dark energy density as $\rho_{\scaleto{DE}{4pt}}$, which is assumed to be constant. Based on these simple assumptions, an estimate for the current amount of dark energy (in the observable universe) may be given by:
\begin{equation}
    E_{\scaleto{DE}{4pt}} = \frac{4\pi}{3}\,r_{\text{obs}}^{3}\,\rho_{\scaleto{DE}{4pt}}\,.\label{Amount-of-DE}
\end{equation}
Now, let us consider the contribution of quintessence and its energy density when the quintessential fluid is the only source of dark energy. To integrate this energy density throughout the whole observable universe, we define:
\begin{equation}
    E_{\scaleto{Q}{4pt}} = \int_{V_{\text{obs}}} \rho_{\scaleto{Q}{4pt}}(r)\, d V = \left(2\, \alpha_{\scaleto{Q}{4pt}} + 1\right) \frac{N_{\scaleto{Q}{4pt}}}{2 G/c^{4}}\,\int_{0}^{r_{\text{obs}}} r^{2\alpha_{\scaleto{Q}{4pt}}}\, dr\,.
\end{equation}
For this type of estimate, it is sufficient to consider $dV = r^{2}\,\sin\theta\,dr\, d\theta\, d\varphi$, in accordance with \cref{metric}. The above definition yields:
\begin{equation}
    E_{\scaleto{Q}{4pt}} = \frac{N_{\scaleto{Q}{4pt}}}{2 G/c^{4}}\, r_{\text{obs}}^{2 \alpha_{\scaleto{Q}{4pt}} + 1}\,,
\end{equation}
which is an approximation of the energy contribution from quintessence as a function of $\alpha_{\scaleto{Q}{4pt}}$ and $r_{\text{obs}}$.

Nevertheless, we can require $E_{\scaleto{Q}{4pt}} = E_{\scaleto{DE}{4pt}}$ to investigate the behavior and orders of magnitude of $N_{\scaleto{Q}{4pt}}$ for different values of $\alpha_{\scaleto{Q}{4pt}}$. By doing so, we obtain that:
\begin{equation}
    N_{\scaleto{Q}{4pt}} = \frac{8\pi G}{3\, c^{4}}\,\rho_{\scaleto{DE}{4pt}}\,r_{\text{obs}}^{2(1-\alpha_{\scaleto{Q}{4pt}})}\,,\label{N_Q_Estimate_Symbolic}
\end{equation}
where $\rho_{\scaleto{DE}{4pt}}$ and $r_{\text{obs}}$ are fixed values. Considering the current radius of the observable Universe to be on the order of $r_{\text{obs}} \approx 4.4 \times 10^{26}$ m \cite{ParticleDataGroup2024,PlanckCosmology2014} and Dark Energy's energy density to be on the order of $\rho_{\scaleto{DE}{4pt}} \approx 6 \times 10^{-10}$ J/m$^{3}$ \cite{ParticleDataGroup2024,PlanckCosmology2014}, we can numerically estimate $N_{\scaleto{Q}{4pt}}$ as:
\begin{equation}
    N_{\scaleto{Q}{4pt}} = \left(4.15 \times 10^{-53}\right) r_{\text{obs}}^{2-2\alpha_{\scaleto{Q}{4pt}}}\, \text{m}^{-2},\label{N_Q_Estimate}
\end{equation}
where we used that $G/c^{4} = 8.24 \times 10^{-45}\text{ kg}^{-1}\text{ m}^{-1}\text{ s}^{2}$ \cite{ParticleDataGroup2024}. Given that we are working in SI units, a dimensional analysis shows that $N_{\scaleto{Q}{4pt}}$ has dimensions of $\text{m}^{-2\alpha_{\scaleto{Q}{3pt}}}$.
\begin{figure}[ht!]
    \centering
    \includegraphics[width=0.65\linewidth]{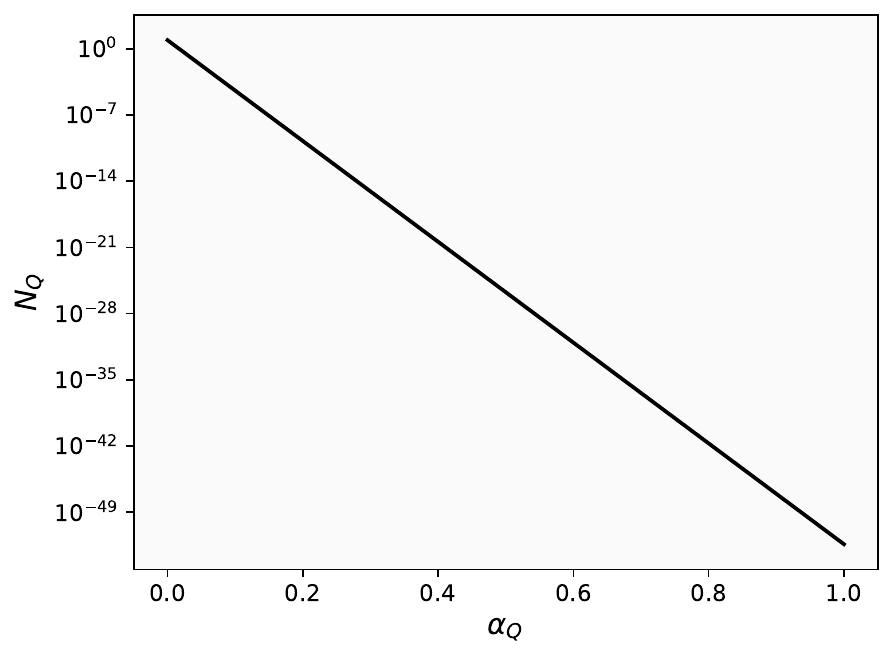}
    \caption{Plot of the quintessence parameter, $N_{\scaleto{Q}{4pt}}$, as a function of the state parameter $\alpha_{\scaleto{Q}{4pt}}$. As shown in \cref{N_Q_Estimate}, $N_{\scaleto{Q}{4pt}}$ exponentially decreases as $\alpha_{\scaleto{Q}{4pt}}$ increases.}
    \label{N_Q_function_Alpha_Q}
\end{figure}

Finally, we present numerical estimates for $N_{\scaleto{Q}{4pt}}$ as a function of $\alpha_{\scaleto{Q}{4pt}}$ in \cref{N_Q_function_Alpha_Q}. We note that, as $\alpha_{\scaleto{Q}{4pt}} \to 1$, the quintessence parameter $N_{\scaleto{Q}{4pt}}$ approaches $4.15 \times 10^{-53}$ m$^{-2}$, in accordance with \cref{N_Q_Estimate}.

\subsection{Contribution to the metric}

Now that we have an estimate for the quintessence parameter $N_{\scaleto{Q}{4pt}}$, we must investigate the contribution of $N_{\scaleto{Q}{4pt}}\, r^{2 \alpha_{\scaleto{Q}{4pt}}}$ to the metric function $f(r)$. From \cref{N_Q_Estimate_Symbolic}, it is straightforward to show that:
\begin{equation}
    N_{\scaleto{Q}{4pt}}\, r^{2 \alpha_{\scaleto{Q}{4pt}}} = \frac{2 G}{c^{4}}\,\frac{E_{\scaleto{DE}{4pt}}}{r_{\text{obs}}}\,\left(\frac{r}{r_{\text{obs}}}\right)^{2\alpha_{\scaleto{Q}{4pt}}}\,,\label{Eq_Metric_Contribution}
\end{equation}
implying that the contribution from quintessence to the metric becomes significant only when $r$ is comparable to $r_{\text{obs}}$, as long as $\alpha_{\scaleto{Q}{4pt}}>0$. The maximum contribution occurs in the limiting case where $\alpha_{\scaleto{Q}{4pt}} = 0$; as $\alpha_{\scaleto{Q}{4pt}}$ increases, the quintessence contribution decreases exponentially. 

\begin{figure}[!ht]
    \centering
    \includegraphics[width=0.6\linewidth]{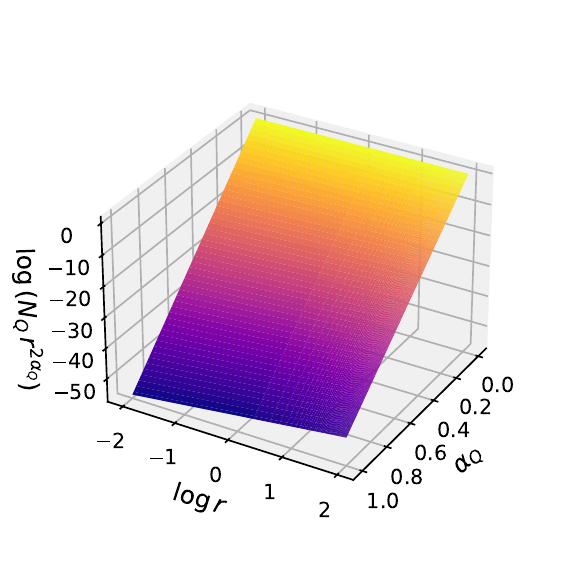} \caption{Contribution of the quintessential term on the metric function $f(r)$. As represented in \cref{Contribution_Numeric_Estimate}, the influence of the quintessence term is most significant at smaller values of $\alpha_{\scaleto{Q}{4pt}}$. In contrast, as the state parameter approaches its upper bound, significantly larger radii are required for the quintessence effect to become substantial.}
    \label{N_Q-term-contribution}
\end{figure}

\begin{figure}[ht!]
    \centering    
    \includegraphics[width=0.65\linewidth]{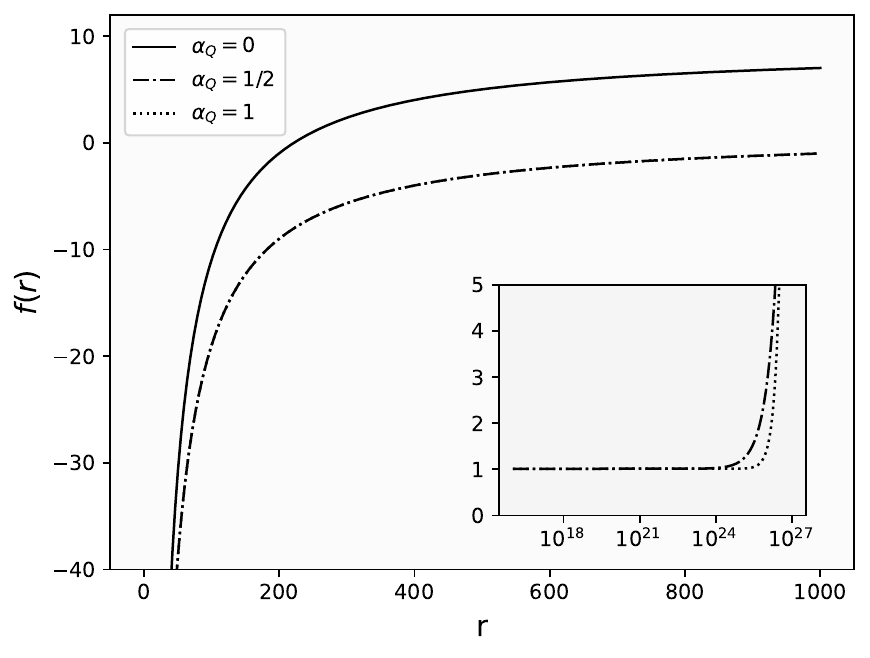}
    \caption{Behavior of the metric function for the state parameter $\alpha_{\scaleto{Q}{4pt}}=0,\, 1/2,\, 1$. The cloud contribution and the Schwarzschild radius are set as $\overline{a}=10^{-2}$ and $r_{s}=2 \times 10^{3}$ m, respectively, while the quintessential term is represented according to \cref{Contribution_Numeric_Estimate}, where $r_{\text{obs}} = 4.4 \times 10^{26}$ m. As highlighted in the inset plot, the curves concerning the two higher state parameters will substantially distinguish themselves near $r \propto r_{\text{obs}}$.}
    \label{full-f(r)-all-alpha}
\end{figure}

The total amount of dark energy, $E_{\scaleto{DE}{4pt}} \approx 2.14 \times 10^{71}$ J, was calculated using \cref{Amount-of-DE} with the values previously specified for $\rho_{\scaleto{DE}{4pt}}$ and $r_{\text{obs}}$. Numerically, \cref{Eq_Metric_Contribution} can be expressed as:
\begin{equation}
    N_{\scaleto{Q}{4pt}} r^{2 \alpha_{\scaleto{Q}{4pt}}} \approx 8.02 \left(\frac{r}{r_{\text{obs}}}\right)^{2\alpha_{\scaleto{Q}{4pt}}}\,.\label{Contribution_Numeric_Estimate}
\end{equation}
This is illustrated in \cref{N_Q-term-contribution}. As expected, $N_{\scaleto{Q}{4pt}} r^{2 \alpha_{\scaleto{Q}{4pt}}}$ is dimensionless.
For completeness, we display in \cref{full-f(r)-all-alpha} the behavior of the complete metric function for three different values of $\alpha_{\scaleto{Q}{4pt}}$, namely $\alpha_{\scaleto{Q}{4pt}} = 0, \, 1/2, \, 1$. As anticipated, the quintessence parameter is highly sensitive to changes in the state parameter, yielding the most substantial contribution at its lowest value, i.e., $\alpha_{\scaleto{Q}{4pt}}=0$. Furthermore, while the cloud parameter induces a vertical shift in the metric function, the quintessential term enables horizon formation at considerably smaller distances for the lower bound of $\alpha_{\scaleto{Q}{4pt}}$ compared to its higher values. In addition, between $\alpha_{\scaleto{Q}{4pt}}=1/2, \, 1$, the most prominent difference will take place only for $r \propto r_{\text{obs}}$, as shown in the inset plot of \cref{full-f(r)-all-alpha}.

\begin{figure}[ht!]
    \centering
    \minipage{0.49\textwidth}
        \includegraphics[width=\linewidth]{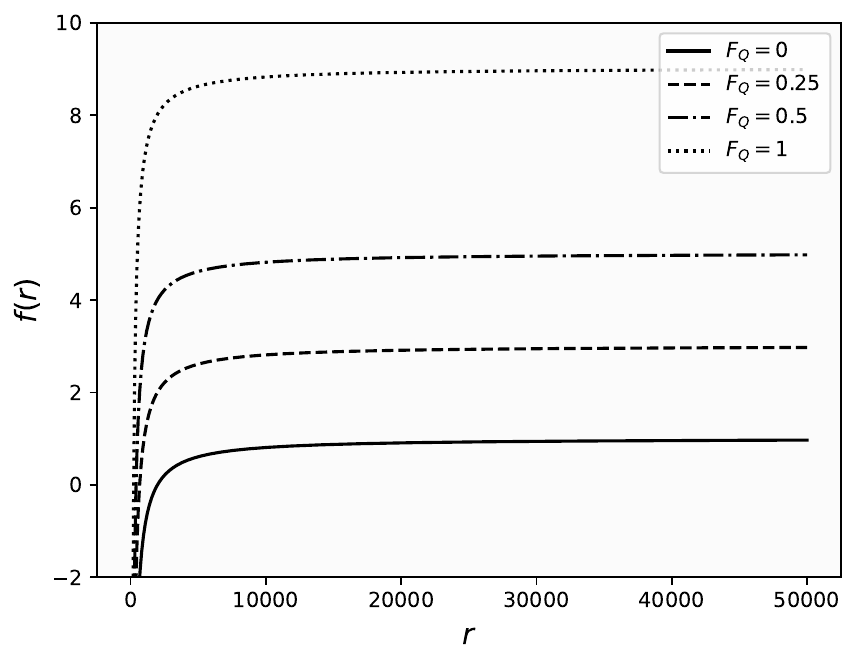}
    \endminipage\hfill
    \minipage{0.49\textwidth}%
        \includegraphics[width=\linewidth]{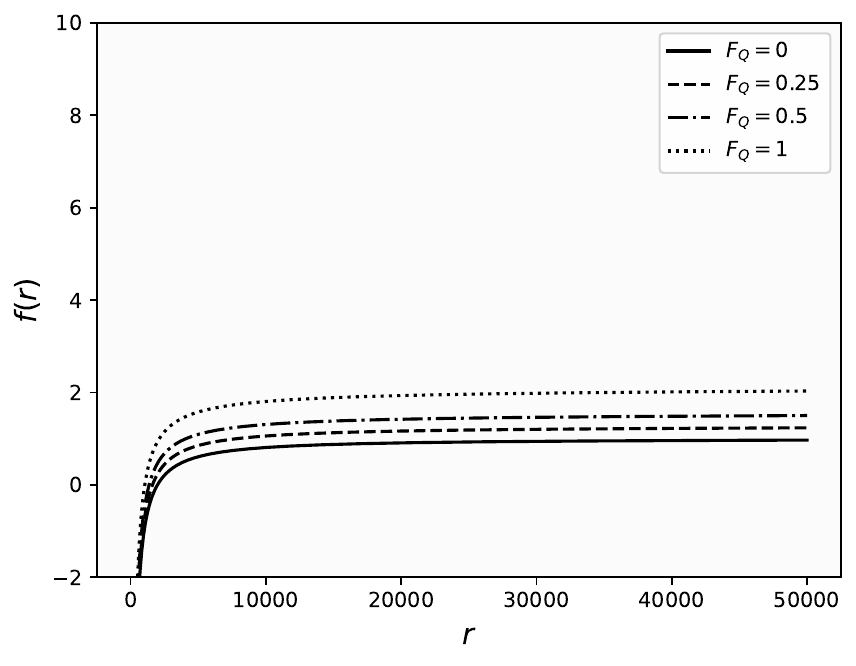}
    \endminipage
    \caption{$f(r)$ for different fractions of quintessence. We set the cloud parameter $\overline{a}=10^{-2}$, the Schwarzschild radius $r_{s}=2 \times 10^{3}$ m, and the dark term according to \cref{fraction_NQ_term}.     
    The left panel displays the metric function for $\alpha_{\scaleto{Q}{4pt}} = 0$, while the right panel illustrates $f(r)$ for $\alpha_{\scaleto{Q}{4pt}} = 0.02$.}
    \label{f_of_r_fractions}
\end{figure}

Finally, we consider that quintessence corresponds only partially to the total dark energy content of the universe. We achieve this by treating the dark term of the metric function as
    \begin{equation}
        N_{\scaleto{Q}{4pt}} r^{2 \alpha_{\scaleto{Q}{4pt}}} = 8.02 \left(\frac{r}{r_{\text{obs}}}\right)^{2\alpha_{\scaleto{Q}{4pt}}}\,  F_{\scaleto{Q}{4pt}}\, , 
        \label{fraction_NQ_term}
    \end{equation}
with $F_{\scaleto{Q}{4pt}}$ representing the fraction of dark energy attributed to quintessence. 
\Cref{f_of_r_fractions} illustrates the behavior of the metric function for different values of $F_{\scaleto{Q}{4pt}}$, considering $\alpha_{\scaleto{Q}{4pt}}=0$ and $\alpha_{\scaleto{Q}{4pt}}=0.02$.

\begin{figure}[ht!]
    \centering
    \minipage{0.49\textwidth}
        \includegraphics[width=\linewidth]{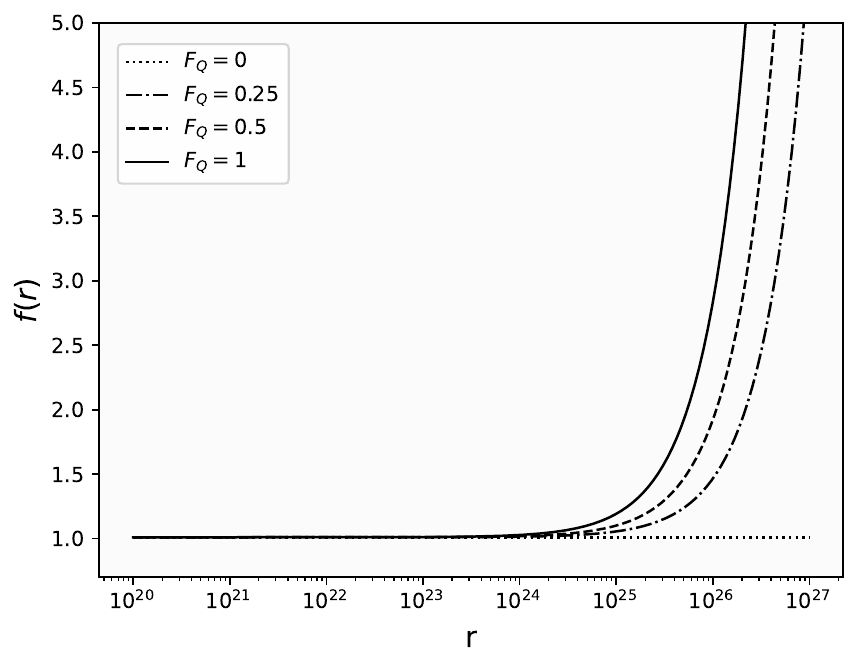}
    \endminipage\hfill
    \minipage{0.49\textwidth}%
        \includegraphics[width=\linewidth]{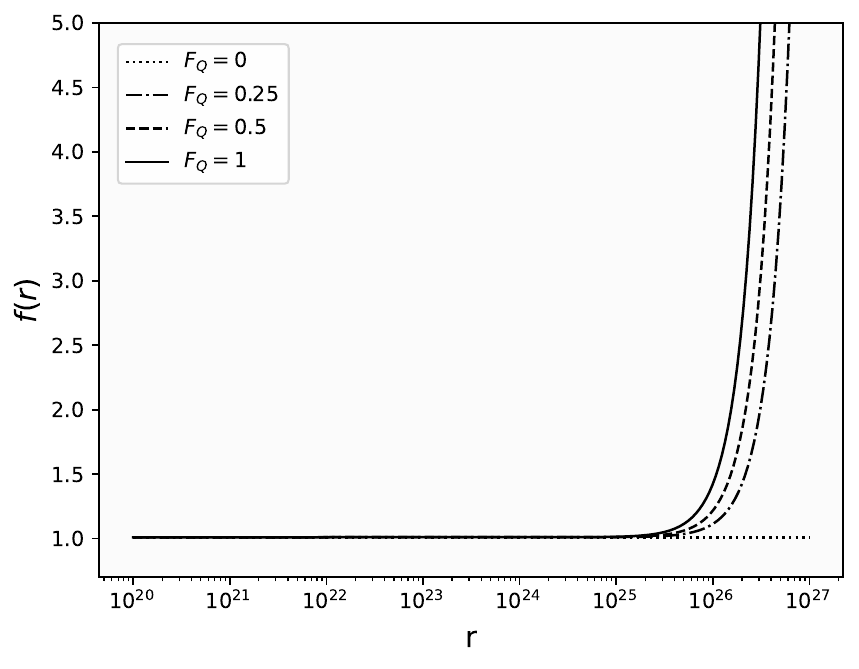}
    \endminipage
    \caption{Behavior of the metric function for different quintessence fractions. The cloud parameter was set to $\overline{a}=10^{-2}$, the Schwarzschild radius is $r_{s}=2 \times 10^{3}$ m, and the dark term was considered according to \cref{fraction_NQ_term}.     
    The left panel displays the metric function for $\alpha_{\scaleto{Q}{4pt}} = 1/2$, while the right panel depicts $f(r)$ for $\alpha_{\scaleto{Q}{4pt}} = 1$.}
    \label{f_of_r_fractions_higher_alpha}
\end{figure}

As shown in the left and right panels, a small increase in the state parameter brings the curves closer together. This is because $N_{\scaleto{Q}{4pt}}$ decreases as $\alpha_{\scaleto{Q}{4pt}}$ increases (as depicted in \cref{N_Q_function_Alpha_Q}), leading to even smaller values for its fractions and reducing their macroscopic distinction. Moreover, because the dark term contribution is orders of magnitude smaller than other metric contributions at higher $\alpha_{\scaleto{Q}{4pt}}$, the role of quintessence fractions becomes prominent only at $r \propto r_{\text{obs}}$. \Cref{f_of_r_fractions_higher_alpha} illustrates this feature for $\alpha_{\scaleto{Q}{4pt}}=1/2, \, 1$.

\section{Event Horizon formation} \label{Event_Horizon_Formation}

The existence of event horizons is a fundamental aspect of our analysis. Beyond determining the spacetime structure and, consequently, the behavior of physical particles, the presence of an event horizon provides essential information for solving the Klein-Gordon equation. Therefore, this section investigates this aspect.

Based on the metric function obtained in \cref{Metric_Function}, we observe that each value of the state parameter leads to a unique spacetime behavior, resulting in distinct event horizon radii. To address this, we recall that the existence of any event horizon is dependent on the existence of real positive roots of the equation
\begin{equation}
    f(r_{\scaleto{+}{5pt}}) = 1 + \overline{a} -\frac{r_{s}}{r_{\scaleto{+}{5pt}}} + N_{\scaleto{Q}{4pt}}\, r_{\scaleto{+}{5pt}}^{2 \alpha_{\scaleto{Q}{4pt}}} = 0\,.\label{Event_Horizon_Condition}
\end{equation}
Since the state parameter $\alpha_{\scaleto{Q}{4pt}}$ is continuous, we consider specific values in the interval $(0,\,1)$, according to which the solutions to \cref{Event_Horizon_Condition} are classified.

\begin{figure}[ht!]
    \centering
    \minipage{0.49\textwidth}
        \includegraphics[width=\linewidth, trim={0 {15pt} 0 {50pt}},clip]{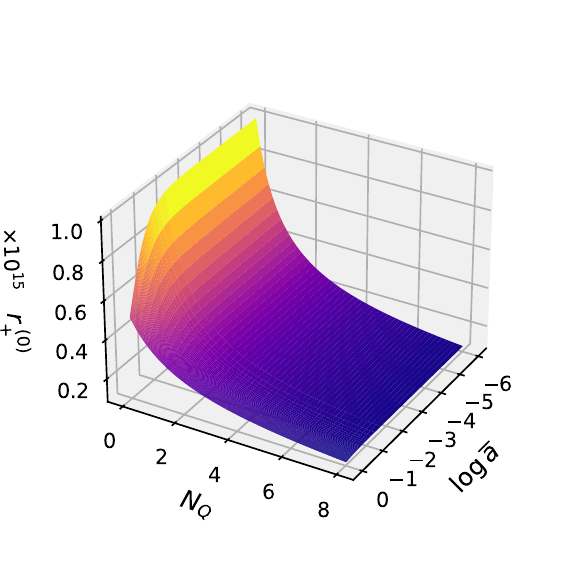}
    \endminipage\hfill
    \minipage{0.49\textwidth}%
        \includegraphics[width=\linewidth, trim={0 {15pt} 0 {50pt}},clip]{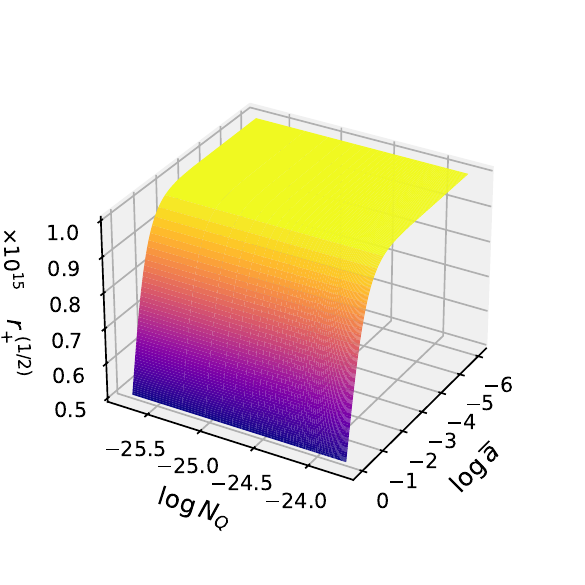}
    \endminipage
    \caption{Horizon radii $r^{\scaleto{(0)}{6pt}}_{+}$ and $r^{\scaleto{(1/2)}{6pt}}_{+}$ as a function of the quintessence and cloud parameters. The Schwarzschild radius is set to $r_{s}=10^{15}$ m. The left panel shows the horizon when $\alpha_{\scaleto{Q}{4pt}}=0$, as described by \cref{Event-Horizon-Alpha_Q-0}, while the right panel depicts the horizon when $\alpha_{\scaleto{Q}{4pt}}=1/2$, according to \cref{Event_Horizon_Meio}.}
    \label{rHorizon_0_Meio}
\end{figure}

We start with the lower bound of the state parameter ($\alpha_{\scaleto{Q}{4pt}} = 0$), where \cref{Event_Horizon_Condition} implies that:
\begin{equation}
    r^{\scaleto{(0)}{6pt}}_{+} = \frac{r_{s}}{1 + \overline{a} + N_{\scaleto{Q}{4pt}}}\,.\label{Event-Horizon-Alpha_Q-0}
\end{equation}
As it is evident from \cref{Event-Horizon-Alpha_Q-0}, the position of the horizon is directly proportional to the Schwarzschild radius, while $\overline{a}$ and $N_{\scaleto{Q}{4pt}}$ equally contribute to regulating $r^{\scaleto{(0)}{6pt}}_{+}$. This behavior is depicted in the left panel of \cref{rHorizon_0_Meio}. Notably, as the quintessence parameter decreases from its estimated value in \cref{N_Q_Estimate}, the horizon radius increases with smaller $\overline{a}$. Moreover, because $\overline{a}$ and $N_{\scaleto{Q}{4pt}}$ similarly contribute in the denominator of \cref{Event-Horizon-Alpha_Q-0}, the cloud parameter has little effect on the horizon's position when $N_{\scaleto{Q}{4pt}}$ is high.

On the other hand, when setting $\alpha\scaleto{Q}{4pt} = 1/2$ in \cref{Event_Horizon_Condition}, one sees that, at the middle of the interval, $f(r_{\scaleto{+}{5pt}})$ satisfies
\begin{equation*}
    1 + \overline{a} -\frac{r_{s}}{r_{\scaleto{+}{5pt}}} + N_{\scaleto{Q}{4pt}}\, r_{\scaleto{+}{5pt}} = 0\,,
\end{equation*}
which leads to
\begin{equation}
    r^{\scaleto{(1/2)}{6pt}}_{+} = \frac{\left(1+\overline{a}\right)}{2\,N_{\scaleto{Q}{4pt}}} \left[\sqrt{1 + \frac{4 N_{\scaleto{Q}{4pt}}\, r_{s}}{\left(1+\overline{a}\right)^{2}}} - 1\right]\,.\label{Event_Horizon_Meio}
\end{equation}
The right panel of \cref{rHorizon_0_Meio} illustrates \cref{Event_Horizon_Meio}. Given that the quintessence parameter is exceptionally small for $\alpha_{\scaleto{Q}{4pt}}=1/2$, the cloud parameter has a more pronounced effect compared to its influence when $\alpha_{\scaleto{Q}{4pt}}=0$. For $N_{\scaleto{Q}{4pt}}$ at least two orders of magnitude above the estimate in \cref{N_Q_Estimate}, decreasing $\overline{a}$ leads to an increase in the horizon's position.

\begin{figure}[ht!]
    \centering
    \includegraphics[width=0.65\linewidth]{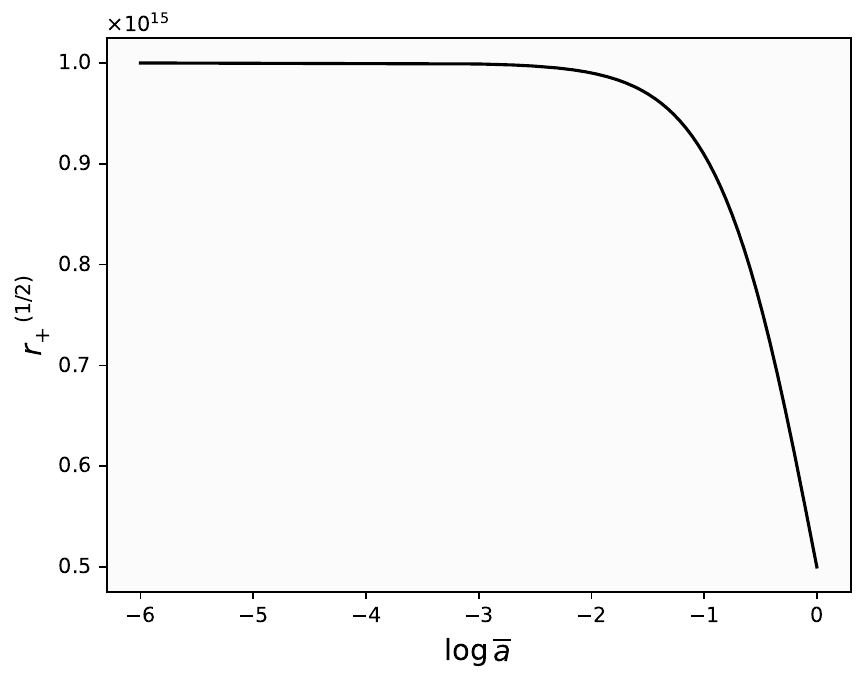}
    \caption{Behavior of \cref{EH-without-quintessence} as a function of the cloud parameter $\overline{a}$. We set the Schwarzschild radius to $r_{s}=10^{15}$ m. Note that \cref{EH-without-quintessence} closely approximates \cref{Event_Horizon_Meio}, which is shown in the right panel of \cref{rHorizon_0_Meio}.}
    \label{rHorizon_approx}
\end{figure}

Furthermore, when $N_{\scaleto{Q}{4pt}}\, r_{s}/(1+\overline{a})^{2} \ll 1$, which is a plausible physical regime\footnote{Remember that, for $\alpha_{\scaleto{Q}{4pt}} = 1/2$, the quintessence parameter $N_{\scaleto{Q}{4pt}}$ is on the order of $10^{-26}$ m$^{-1}$. Hence, even if we consider $\overline{a} = 0$ along with $r_{s} = 10^{15}$ m, the product $N_{\scaleto{Q}{4pt}}\, r_{s}/(1+\overline{a})^{2} \approx 10^{-11}$, which readily satisfies the considered regime.} according to \cref{section-Estimates}, the event horizon position is well approximated by:
\begin{equation}
    r^{\scaleto{(1/2)}{6pt}}_{+} = \frac{r_{s}}{\left(1+\overline{a}\right)}\,.\label{EH-without-quintessence}
\end{equation}
The above result, depicted in \cref{rHorizon_approx}, is also the exact solution for when the quintessential fluid is absent, as implied by \cref{Event-Horizon-Alpha_Q-0}. This was expected, because the contribution from the quintessence term is significantly suppressed by the small $N_{\scaleto{Q}{4pt}} \sim 10^{-26}$ m$^{-1}$ (for distances $r\ll r_{\scaleto{obs.}{4pt}}$). As a result, the Schwarzschild radius becomes the dominant factor in horizon formation, which is evident from \cref{EH-without-quintessence}. A similar effect is anticipated for the upper bound of the state parameter.

\begin{figure}[ht!]
    \centering
    \includegraphics[width=0.6\linewidth, trim={0 {15pt} 0 {50pt}},clip]{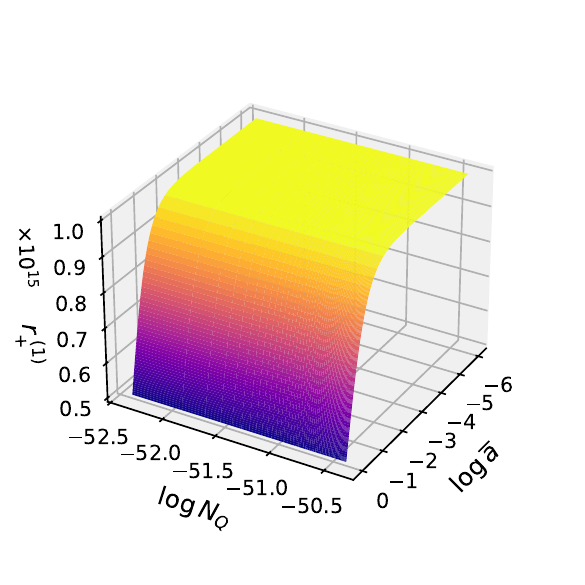}
    \caption{Behavior of \cref{Event_Horizon_Alpha_Q_1_RealRoot} for different quintessence and cloud parameters.  The Schwarzschild radius was set as $r_{s} = 10^{15}$ m. Note that, for the $N_{\scaleto{Q}{4pt}}$ estimate given by \cref{N_Q_Estimate}, the position of the horizon when $\alpha_{\scaleto{Q}{4pt}}=1$ increases with smaller $\overline{a}$.}
    \label{rHorizon_1}
\end{figure}

When $\alpha_{\scaleto{Q}{4pt}} = 1$, the condition $f(r_{\scaleto{+}{5pt}}) = 0$ implies that:
\begin{equation*}
    1 + \overline{a} -\frac{r_{s}}{r_{\scaleto{+}{5pt}}} + N_{\scaleto{Q}{4pt}}\, r_{\scaleto{+}{5pt}}^{2} = 0\,.
\end{equation*}
This equation admits only one real root, given by:
\begin{equation}
    r^{\scaleto{(1)}{6pt}}_{+} = \Delta^{1/3} - \frac{\left(1+\overline{a}\right)}{3\, N_{\scaleto{Q}{4pt}}}\,\Delta^{-1/3}\,,\label{Event_Horizon_Alpha_Q_1_RealRoot}
\end{equation}
with
\begin{equation*}
\Delta = \left(\frac{1+\overline{a}}{3\,N_{\scaleto{Q}{4pt}}}\right)^{3/2}
\left[\sqrt{1 + \frac{27}{4}\,\frac{N_{\scaleto{Q}{4pt}}\,r_{s}^{2}}{\left(1 + \overline{a}\right)^{3}}} + \sqrt{\frac{27}{4}\,\frac{N_{\scaleto{Q}{4pt}}\,r_{s}^{2}}{\left(1+\overline{a}\right)^{3}}}\,\right]\,.
\end{equation*}
In a previous work on black strings \cite{Deglmann2025}, we showed that an equivalent equation for $r^{\scaleto{(1)}{6pt}}_{+}$ could be approximated to the form of \eqref{EH-without-quintessence} provided that $N_{\scaleto{Q}{4pt}}\,r_{s}^{2}/\left(1+\overline{a}\right)^{3} \ll 1$. We shall retain this result for future convenience, given that the parameter $N_{\scaleto{Q}{4pt}}$ is expected to be on the order of $10^{-52}$ m$^{-2}$. Then, even under the best hypothesis for $\overline{a}$ and $r_{s}$ (to maximize the overall product), $N_{\scaleto{Q}{4pt}}\,r_{\scaleto{S}{4pt}}^{2}/\left(1+\overline{a}\right)^{3}$ will be on the order of $10^{-22}$. For completeness, \cref{rHorizon_1} displays $r^{\scaleto{(1)}{6pt}}_{+}$ as a function of the quintessence and cloud parameters. Similar to the $\alpha_{\scaleto{Q}{4pt}}=1/2$ case, high values of $\overline{a}$ lead to a notable increase in the horizon position. As this holds true for increasing orders of $N_{\scaleto{Q}{4pt}}$, \cref{rHorizon_approx} further clarifies that \cref{EH-without-quintessence} remains a close approximation within the regime where $N_{\scaleto{Q}{4pt}}\,r_{s}^{2}/\left(1+\overline{a}\right)^{3} \ll 1$.

Nevertheless, we highlight that as the quintessential term diminishes with an increasing state parameter (see \cref{N_Q_function_Alpha_Q}), suggesting its suppression by the Schwarzschild radius for higher $\alpha_{\scaleto{Q}{4pt}}$, this effect is limited to the macroscopic description of the horizon. Crucially, it does not signify negligibility at the quantum level. Despite \cref{EH-without-quintessence} representing an approximation for the horizon radius with $\alpha_{\scaleto{Q}{4pt}}=1/2$ and $1$, the quintessential term $N_{\scaleto{Q}{4pt}}$ arises from distinct elements within the wave function. We shall explore this in the following sections.

\section{The Klein-Gordon equation in the Schwarzschild-Kiselev-Letelier spacetime}\label{KG-equation}

Having analyzed the quintessence contribution to the Schwarzschild-Kiselev-Letelier spacetime and its impact on horizon formation, we now turn our attention to the quantum effects that quintessence may exert on the matter content of this background. To address this subject, we investigate the behavior of scalar particles embedded in this spacetime to obtain solutions near the event horizon. These solutions, as will be explored in \cref{Section_DPs}, exhibit a phase difference due to the quintessence's effect on the spin$-0$ particle.

For the moment, we proceed with the Klein-Gordon equation in curved spacetime with the mostly-minus sign convention, that is
\begin{equation}
    \frac{1}{\sqrt{-g}}\,\partial_{\mu}\left(g^{\mu\nu}\sqrt{-g}\,\partial_{\nu}\,\phi\right) + \frac{m_{\phi}^{2}\,c^{2}}{\hbar^{2}}\,\phi = 0\,,
    \label{Klein-Gordon-equation}
\end{equation}
where $g=\text{det}(g_{\mu\nu})$ and $m_{\phi}$ is the spin$-0$ particle's mass. In addition, the metric tensor $g_{\mu\nu}$ is fully determined by \cref{Metric_Function}. 

We decouple \cref{Klein-Gordon-equation} using the ansatz $\phi = T(t) \, R(r)\, \Theta(\theta) \, \Phi(\varphi)$, in which the scalar field $\phi$ is now expressed with four independent functions, being
\begin{equation}
        \phi = e^{-i E\,t/\hbar} \, e^{i k \varphi} \, P_{\ell}^{k}(\cos \theta) \, R(r),
\end{equation}
where $E$ corresponds to the particle's energy. Moreover, $k \in \mathbb{Z}$, $\ell \in \mathbb{Z}^{+}$ and $P_{\ell}^{k}(\cos \theta)$ are Legendre functions, encompassing the angular behavior due to the spherical symmetry. The remaining function $R(r)$ is the solution to the radial Klein-Gordon equation, given by:
\begin{equation}
        R''(r) + \left[ \, \frac{f'(r)}{f(r)} + \frac{2}{r} \,  \right]  R'(r) + \left[\frac{\epsilon^{2}}{f(r)^{2}} - \frac{\overline{m}_{\phi}^{2}}{f(r)}  - \frac{\ell(\ell+1)}{r^{2}f(r)} \,  \right] R(r) = 0\,,
        \label{Full-Radial-KG}
\end{equation}
where, here, the prime denotes the derivative with respect to $r$. The parameters $\epsilon$ and $\overline{m}_{\phi}$ are defined as:
\begin{equation}
        \epsilon = \frac{E}{\hbar c}\,,\quad\text{ and }\quad
        \overline{m}_{\phi} = \frac{m_{\phi}\,c}{\hbar}\,.\label{def_Epsilon_M_bar}
\end{equation}
Therefore, both parameters have units of m$^{-1}$ (in SI units).

To handle \cref{Full-Radial-KG} more conveniently, we use the Liouville normal form \cite{Titchmarsh1962} by introducing an auxiliary function $u(r)$,  whereas $R(r)$ is 
\begin{equation}
R(r) = \frac{R_{0}}{r \sqrt{f(r)}} u(r) \, ,\label{R_u_relation}
\end{equation}
with $R_{0}$ being a normalization constant. 

In terms of the auxiliary function, the corresponding radial Klein-Gordon equation becomes:
\begin{equation}
        u''(r) + V_{\text{eff}}(r) \,u(r) = 0 \, ,
        \label{Normal_Radial(r)}
\end{equation}
where $V_{\text{eff}}(r)$ is identified as the effective potential, satisfying:
\begin{equation}
         V_{\text{eff}}(r) = \frac{1}{f(r)^{2}} \left[ \epsilon^{2}  +  \frac{f'(r)^{2}}{4} \right] - \frac{1}{f(r)}  \left[\overline{m}_{\phi}^{2}  + \frac{f''(r)}{2}  + \frac{f'(r)}{r} + \frac{\ell (\ell + 1 ) }{r^{2}} \right] \, .
    \label{V_eff(r)}
\end{equation}
%Tenho que mexer neste parágrafo.
At this point, we introduce a convenient change of variables to incorporate the event horizon into \cref{Metric_Function,Normal_Radial(r),V_eff(r)}. For this purpose, we define the dimensionless coordinate $x = r/r_{\scaleto{+}{5pt}}$. This definition is valid as long as $r_{s}\neq 0$, implying the existence of an event horizon at $r_{\scaleto{+}{5pt}}$. Then, recalling the result of \cref{Metric_Function}, we can rewrite the metric function $f(r)$ in terms of $x$. By doing  so, we obtain that:
\begin{equation}
    f(x) = 1+\overline{a} - \frac{r_{s}}{r_{\scaleto{+}{5pt}}}\,\frac{1}{x} + N_{\scaleto{Q}{4pt}}\,r_{\scaleto{+}{5pt}}^{2\alpha_{\scaleto{Q}{3pt}}}\,x^{2\alpha_{\scaleto{Q}{3pt}}}\,.
\end{equation}
Next, we use the condition $f(r_{\scaleto{+}{5pt}}) = 0$, detailed in \cref{Event_Horizon_Condition}, to show that $f(x)$ is determined by
\begin{equation}
    f(x) = \left(1 + \overline{a}\right)\left(1-\frac{1}{x}\right) + N_{\scaleto{Q}{4pt}}\,r_{\scaleto{+}{5pt}}^{2\alpha_{\scaleto{Q}{3pt}}} \left(x^{2\alpha_{\scaleto{Q}{3pt}}} - \frac{1}{x}\right)\,,\label{f(x)_Part1}
\end{equation}
which is an exact result and holds for every scenario, provided $r_{s}\neq 0$. Under a simple rearrangement, we show that \cref{f(x)_Part1} can be conveniently written as
\begin{equation}
    f(x) = \left(1 + \overline{a} + N_{\scaleto{Q}{4pt}}\,r_{\scaleto{+}{5pt}}^{2\alpha_{\scaleto{Q}{3pt}}}\right)\left(1-\frac{1}{x}\right) + N_{\scaleto{Q}{4pt}}\,r_{\scaleto{+}{5pt}}^{2\alpha_{\scaleto{Q}{3pt}}} \left(x^{2\alpha_{\scaleto{Q}{3pt}}} - 1\right)\,.\label{f(x)_Part2}
\end{equation}
Notably, when $\alpha_{\scaleto{Q}{4pt}}=0$, the term $\left(x^{2\alpha_{\scaleto{Q}{3pt}}} - 1\right)$ becomes null, implying a simpler expression for $f(x)$, which -- as we shall discuss in  \cref{Exact_Solution} -- leads to an exact solution for $R(x)$. On the other hand, to investigate the behavior of the solution for other values of $\alpha_{\scaleto{Q}{4pt}}$, we shall consider a series approximation of $f(x)$ with respect to $x=1$.

Before moving forward, it is necessary to rewrite the auxiliary equation \eqref{Normal_Radial(r)} and the effective potential \eqref{Effective_Potential} in terms of $x$. Since $d^{2}/d r^{2} = (1/r_{\scaleto{+}{5pt}}^{2})\,d^{2}/d x^{2}$, this results in:
\begin{equation}
    u''(x) + r_{\scaleto{+}{5pt}}^{2}\,V_{\text{eff}}(x) \,u(x) = 0\,,
        \label{Normal_u_of_x}
\end{equation}
where the effective potential becomes
\begin{equation}
        r^{2}_{\scaleto{+}{5pt}}\,V_{\text{eff}}(x) = \frac{1}{f(x)^{2}}
        \left[r^{2}_{\scaleto{+}{5pt}}\,\epsilon^{2} + \frac{f'(x)^{2}}{4}\right]
        - \frac{1}{f(x)} 
        \left[\frac{f''(x)}{2} + \frac{f'(x)}{x} + \frac{\ell (\ell +1)}{x^{2}} + r^{2}_{\scaleto{+}{5pt}}\,\overline{m}_{\phi}^{2}\right]\,.\label{Effective_Potential}
\end{equation}
Observe that, in this new coordinate, the event horizon is located at $x=1$, so that the domain for the radial Klein-Gordon solution, $R(x)$, is $x\in(1,\,\infty)$. This means that the particle does not reach the event horizon itself.

\subsection{Exact solution at the lower bound}\label{Exact_Solution}

This subsection presents an exact solution to the Klein-Gordon equation in the limiting case where the quintessence parameter $\alpha_{\scaleto{Q}{4pt}}$ vanishes, corresponding to $\omega_{\scaleto{Q}{4pt}} = -1/3$. Using our previous results of the metric function $f(x)$, the auxiliary differential equation for $u(x)$, and the general expression for the effective potential $r_{\scaleto{+}{5pt}} V_{\text{eff}}(x)$ (please, see \cref{f(x)_Part2,Normal_u_of_x,Effective_Potential}), we show that this solution is exact and can be expressed in terms of the Confluent Heun function\footnote{Here we understand \enquote{Heun functions} as the analytic continuation of the local solutions, which can be determined using the Fuchs-Frobenius method.}. Furthermore, we discuss the impact and limitations of the Frobenius solution to the confluent Heun equation, highlighting the spectral constraint associated with confluent Heun polynomials.

First, let us recall that, when $\alpha_{\scaleto{Q}{4pt}} = 0$, the metric function $f(x)$ from \cref{f(x)_Part2} simplifies to
\begin{equation}
	f(x) = \left(1 + \overline{a} + N_{\scaleto{Q}{4pt}}\right)\left(1-\frac{1}{x}\right)\,.\label{f(x)_Exact_Solution}
\end{equation}
This result is valid for all $x\in \left(1,\,\infty\right)$. By substituting this expression for $f(x)$ into the effective potential \eqref{Effective_Potential}, we obtain that:
\begin{equation}
	\begin{aligned}
		r_{\scaleto{+}{5pt}}^{2}\,V_{\text{eff}}(x) 
		&= \frac{1/4}{x^{2}\left(x-1\right)^{2}} + \frac{r_{s}^{2}\,\epsilon^{2}}{\left(1+\overline{a}+N_{\scaleto{Q}{4pt}}\right)^{4}}\,\frac{x^{2}}{\left(x-1\right)^{2}}\\[5pt] 
		&- \frac{\ell\left(\ell+1\right)}{(1+\overline{a}+N_{\scaleto{Q}{4pt}})}\,
		\frac{1}{x\left(x-1\right)} 
		- \frac{r_{s}^{2}\,\overline{m}_{\phi}^{2}}{\left(1+\overline{a}+N_{\scaleto{Q}{4pt}}\right)^{3}}\,\frac{x}{\left(x-1\right)}\,.
	\end{aligned}\label{Effective_Potential_Exact}
\end{equation}
To map the combined results of the auxiliary differential equation for $u(x)$ and the above effective potential (given by \cref{Normal_u_of_x,Effective_Potential_Exact}) to the general normal form of the confluent Heun equation (as detailed in \cref{Eq_u_Normal_Form}), we decompose \cref{Effective_Potential_Exact} into partial fractions. This decomposition yields:
\begin{equation}
	\begin{aligned}
		r_{\scaleto{+}{5pt}}^{2}\,V_{\text{eff}}(x)
		&= \frac{1/2 + L_{0}}{x}
		- \frac{1/2 - 2 r_{s}^{2}\,\epsilon_{0}^{2} + L_{0} + r_{s}^{2}\,\overline{m}_{0}^{2}}{\left(x-1\right)}
		+ \frac{1/4}{x^{2}}\\[5pt]
		&+ \frac{1/4 + r_{s}^{2}\,\epsilon_{0}^{2}}{\left(x-1\right)^{2}} + r_{s}^{2}\left(\epsilon_{0}^{2} - \overline{m}_{0}^{2}\right)\,,
	\end{aligned}
\end{equation}
where
\begin{equation}
	\begin{aligned}
		L_{0} 
		&= \frac{\ell\left(\ell+1\right)}{\left(1+\overline{a}+N_{\scaleto{Q}{4pt}}\right)}\,,\\
		\epsilon_{0} 
		&= \frac{\epsilon}{\left(1+\overline{a}+N_{\scaleto{Q}{4pt}}\right)^{2}}\,,\\
		\overline{m}_{0} 
		&= \frac{\overline{m}_{\phi}}{\left(1+\overline{a}+N_{\scaleto{Q}{4pt}}\right)^{3/2}}\,.
	\end{aligned}\label{Useful_Definitions}
\end{equation}
Since the quintessence parameter $N_{\scaleto{Q}{4pt}}$ is dimensionless for $\alpha_{\scaleto{Q}{4pt}}=0$ (see \cref{Dimension_N_Q}), we conclude that $L_{0}$ is dimensionless, while both $\epsilon_{0}$ and $\overline{m}_{0}$ have dimensions of m$^{-1}$.

Next, we use the results from \cref{Eq_u_Normal_Form,Parameters_HeunC_Normal,Sol_Geral_u} to show that the exact solution to $u(x)$ is determined by:
\begin{equation}
    \begin{aligned}
	u(x) &= x^{1/2}\,\left(x - 1\right)^{(1 + \gamma)/2}\,e^{\alpha x/2}\,\left[c_{1}\,\text{HeunC}\left(-\alpha,\,\gamma,\,\beta,\,-\delta,\,\delta+\eta;\,x\right)\right.\\
    &\left.+ c_{2}\,\left(x-1\right)^{-\gamma}\,\text{HeunC}\left(-\alpha,\,-\gamma,\,\beta,\,-\delta,\,\delta+\eta;\,x\right)\right]\,,\label{u_EX}
    \end{aligned}
\end{equation}
where the Heun parameters $\alpha$, $\beta$, $\gamma$, $\delta$, and $\eta$ are given by
\begin{equation}
	\begin{aligned}
		\alpha 
		&= \frac{\pm 2 i\,r_{s}}{\left(1+\overline{a}+N_{\scaleto{Q}{4pt}}\right)^{2}}\, \sqrt{\epsilon^{2} - \left(1+\overline{a}+N_{\scaleto{Q}{4pt}}\right) \overline{m}_{\phi}^{2}}\,,\\[5pt]
		\beta 
		&= 0\,,\\[5pt]
		\gamma 
		&= \frac{\pm 2 i\,r_{s}}{\left(1+\overline{a}+N_{\scaleto{Q}{4pt}}\right)^{2}}\,,\\[5pt]
		\delta 
		&= \frac{r_{s}^{2}}{\left(1+\overline{a}+N_{\scaleto{Q}{4pt}}\right)^{4}} \left[2 \epsilon^{2} - \left(1+\overline{a}+N_{\scaleto{Q}{4pt}}\right)\overline{m}_{\phi}^{2}\right]\,,\\[5pt]
		\eta 
		&= -\frac{\ell\left(\ell+1\right)}{\left(1+\overline{a}+N_{\scaleto{Q}{4pt}}\right)}\,,
	\end{aligned}
\end{equation}
and $\text{HeunC}\left(-\alpha,\,\gamma,\,\beta,\,-\delta,\,\delta+\eta;\,x\right)$ must be understood as the analytic continuation of the series solution \eqref{CHE_Fuchs_Frobenius_Solution} with the characteristic exponent $r_{1}=0$. The above results, while expressible in a more compact form using the definitions in \cref{Useful_Definitions}, are intentionally written out in terms of the physical parameters $\overline{a}$, $N_{\scaleto{Q}{4pt}}$, and $\overline{m}_{\phi}$ to highlight this dependence.

Subsequently, by combining the relation \eqref{R_u_relation} with \cref{f(x)_Exact_Solution,u_EX}, we derive the explicit form of the radial Klein-Gordon solution:
\begin{equation}
    \begin{aligned}
	R_{\scaleto{EX}{4pt}}^{\scaleto{(0)}{6pt}}(x) 
    &= \left(x-1\right)^{\gamma/2} e^{\alpha x/2}\,\left[c_{1}\,\text{HeunC}\left(-\alpha,\,\gamma,\,\beta,\,-\delta,\,\delta+\eta;\,x\right)\right.\\
    &\left.+ c_{2}\,\left(x-1\right)^{-\gamma}\,\text{HeunC}\left(-\alpha,\,-\gamma,\,\beta,\,-\delta,\,\delta+\eta;\,x\right)\right]\,,\label{RWF_Exact}
    \end{aligned}
\end{equation}
where $c_{1}$ and $c_{2}$ absorbed the remaining multiplicative constants. The superscript \enquote{$(0)$} denotes the case where $\alpha_{\scaleto{Q}{4pt}} = 0$ and the subscript \enquote{\textsc{ex}} indicates this is an exact solution.

As shown in the appendix, the function  $\text{HeunC}\left(\alpha,\,\beta,\,\gamma,\,\delta,\,\eta;\,x\right)$ can be determined by solving the confluent Heun equation via the Fuchs-Frobenius method. The results originally converge for $1 < x < 2$, due to the regular singularities in the confluent Heun equation. For values of $x \geq 2$, numerical solutions are required. However, the local solution $\text{HeunC}\ell\left(\alpha,\,\beta,\,\gamma,\,\delta,\,\eta;\,x\right)$ is useful when describing a particle near the event horizon. Since $x= r/r_{\scaleto{+}{5pt}}$, any \enquote{distance} is measured with respect to the size of the event horizon radius. For instance, if $r_{\scaleto{+}{5pt}} = 10^{5}$ m, we can use the series solution to describe a particle located between $10^{5}$ m and $2 \times 10^{5}$ m from the origin.

Within this region of convergence (where $1<x<2$), the local confluent Heun function (given by the series \eqref{CHE_Fuchs_Frobenius_Solution}) can be truncated into a $N$--th order polynomial. This requires imposing the $\delta$--condition from \cref{Delta_Condition}, which is a necessary but not sufficient condition for obtaining confluent Heun polynomials. For our exact solution, this constraint takes the form:
\begin{equation}
	\begin{aligned}
		N+1 = \mp \frac{r_{s}}{2\beta_{\scaleto{Q}{5pt}}^{2}} \left[\frac{\beta_{\scaleto{Q}{5pt}}\,\overline{m}_{\phi}^{2} - 2\epsilon^{2}}{\sqrt{\beta_{\scaleto{Q}{5pt}}\,\overline{m}_{\phi}^{2} - \epsilon^{2}}} + 2\right]\,.\label{Delta_Constraint}
	\end{aligned}
\end{equation}
The above result does not provide the full set of eigenvalues $\epsilon$; rather, it yields only those associated with the polynomial solutions. In the next subsection, we investigate an analytical solution to the radial Klein-Gordon equation that is valid for a particle near the event horizon. This solution, based on a first-order approximation, encompasses the general case of $0 < \alpha_{\scaleto{Q}{4pt}} < 1$ and reveals the emergence of what we refer to as \enquote{dark phases}.

\subsection{Solution near the event horizon}

While the previous section provided an exact solution for the specific case of $\alpha_{\scaleto{Q}{4pt}}=0$, a more general approach is required to understand the system's behavior for $0<\alpha_{\scaleto{Q}{4pt}}<1$. In this subsection, we present a first-order analytical solution to the Klein-Gordon equation by restricting the analysis to a small neighborhood of the event horizon (characterized by $x\to 1^{+}$). We start by deriving a linear approximation of the metric function $f(x)$ through a series expansion around $x=1$. This simplified metric, in turn, allows us to express the radial Klein-Gordon solution in terms of the confluent Heun functions. Then, we investigate the behavior of this formal solution in the near-horizon limit to ensure its compatibility with the restricted domain. These results will be used in the next section to investigate the emergence of \enquote{dark phases}.

A general series expansion of $f(x)$ (given by \cref{f(x)_Part2}) around the event horizon (at $x=1$) is:
\begin{equation}
	f(x) = \left(1 + \overline{a} + N_{\scaleto{Q}{4pt}}\, r_{\scaleto{+}{5pt}}^{2\alpha_{\scaleto{Q}{3pt}}}\right)\sum_{n=1}^{+\infty}\left(-1\right)^{n+1}\left(x-1\right)^{n}
	\,+\, N_{\scaleto{Q}{4pt}}\, r_{\scaleto{+}{5pt}}^{2\alpha_{\scaleto{Q}{3pt}}}
	\sum_{n=1}^{+\infty} \binom{2\alpha_{\scaleto{Q}{4pt}}}{n} \left(x-1\right)^{n}\,.\label{series-expansion}
\end{equation}
The radius of convergence for both series is $|x-1| < 1$, as the first corresponds to the expansion of $(1 - 1/x)$ and the second is a binomial series for $x^{2\alpha_{\scaleto{Q}{3pt}}} = \left(1+\left(x-1\right)\right)^{2\alpha_{\scaleto{Q}{3pt}}}$.

By restricting the particle's domain to a small neighborhood of the event horizon, formally $x\to 1^{+}$, we can show that the first-order terms of the series expansion \cref{series-expansion} provide the dominant contribution to $f(x)$.

To illustrate this, consider a particle in the range $1.001 \leq x \leq 1.01$. In this case, the first-order term $\left(x-1\right)$ is in the range $10^{-3}$ to $10^{-2}$ while the second-order term $\left(x-1\right)^{2}$ ranges from $10^{-6}$ to $10^{-4}$. As higher-order terms are even more negligible, we can reasonably approximate $f(x)$ as:
\begin{equation}
	f(x) = \beta_{\scaleto{+}{5pt}} (x-1),\label{Metric_Function_NH}
\end{equation}
where
\begin{equation}
	\beta_{\scaleto{+}{5pt}} = 1 + \overline{a} + \left(2\alpha_{\scaleto{Q}{4pt}} + 1\right) N_{\scaleto{Q}{4pt}}\,r_{\scaleto{+}{5pt}}^{2\alpha_{\scaleto{Q}{3pt}}}\,.\label{def_Beta_Plus}
\end{equation}
We emphasize that $\beta_{\scaleto{+}{5pt}}$, defined by \cref{def_Beta_Plus}, is simply $\beta_{\scaleto{+}{5pt}} = r_{\scaleto{+}{5pt}} [df(r)/dr]|_{r_{\scaleto{+}{5pt}}}$, in accordance with the series expansion \eqref{series-expansion} for $f(x)$. Furthermore, we stress that the first order approximation \eqref{Metric_Function_NH} holds only for $x\to 1^{+}$. For other regions, higher-order corrections would be mandatory.

Before proceeding, we must address two important considerations regarding our near-horizon approximation. 
First, it is important to note that a small domain for the dimensionless coordinate $x$ does not necessarily imply a small physical region. Due to the relation $r = x r_{\scaleto{+}{5pt}}$, a large event horizon radius can lead to significant distances (away from the horizon itself). For example, if $r_{\scaleto{+}{5pt}} = 10^{8}$ m, an interval of $\Delta x = 10^{-3}$ corresponds to a radial distance of $10^{5}$ m.
Second, our approximation is restricted to a first-order term because, while the general effective potential is already non-linear, including higher-order terms from the series expansion of $f(x)$ would introduce additional nonlinearities that prevent us from obtaining an analytical solution.

Now that the first-order approximation of the metric function has been established, we can substitute \cref{Metric_Function_NH} into \cref{Effective_Potential}. A partial fraction decomposition of the resulting expression for the effective potential yields:
\begin{equation}
    \begin{aligned}
          r^{2}_{\scaleto{+}{5pt}}\,V_{\text{eff}}(x) &= \frac{1}{x} \left[1 + \frac{\ell(\ell+1)}{\beta_{\scaleto{+}{5pt}}} \right] + \frac{1}{(x-1)} \left[ - \frac{\overline{m}_{\phi}^{2} r^{2}_{\scaleto{+}{5pt}}}{\beta_{\scaleto{+}{5pt}}} - \frac{\ell(\ell+1)}{\beta_{\scaleto{+}{5pt}}} -1 \right]  \\[5pt] & + \frac{1}{x^{2}} \left[ \frac{\ell(\ell+1)}{\beta_{\scaleto{+}{5pt}}} \right] + \frac{1}{(x-1)^{2}} \left[  \frac{\epsilon^{2} r^{2}_{\scaleto{+}{5pt}}}{\beta^{2}_{\scaleto{+}{5pt}}} + \frac{1}{4}\right] \, . 
    \end{aligned}
    \label{V_eff(x)_final}
\end{equation}
By comparing \cref{Normal_u_of_x,V_eff(x)_final} with the general normal form of the Confluent Heun equation (as detailed in the appendix), we find that the solution for $u(x)$ is:
\begin{equation}
	\begin{aligned}
        u(x) 
        &= \exp{(\alpha x/2)}\, x^{(\beta+1)/2} (x-1)^{(\gamma +1)/2} \left[c_{1}\,\text{HeunC}\left(-\alpha,\,\gamma,\,\beta,\,-\delta,\,\delta+\eta;\,x\right) \right.  \\[5pt] 
        &+ \left. c_{2}\, \left(x-1\right)^{-\gamma}\,\text{HeunC}\left(-\alpha,\,-\gamma,\,\beta,\,-\delta,\,\delta+\eta;\,x\right)\right]\,.
	\end{aligned}\label{u_of_x_NH}
\end{equation}
From the comparison with \cref{Parameters_HeunC_Normal}, the Heun parameters $\alpha$, $\beta$, $\gamma$, $\delta$, and $\eta$ are found to be:
\begin{equation}
    \begin{aligned}
        \alpha &= 0\,,\\[5pt]
        \beta  &= \pm 2 i\,\left[\frac{\ell (\ell +1)}{\beta_{\scaleto{+}{5pt}}} - \frac{1}{4}\right]^{1/2}\,,\\[5pt]
        \gamma &= \pm 2 i\,\frac{\epsilon\,r_{\scaleto{+}{5pt}}}{\beta_{\scaleto{+}{5pt}}}\,,\\[5pt]
        \delta &= - \frac{\overline{m}_{\phi}^{2}\, r^{2}_{\scaleto{+}{5pt}}}{\beta_{\scaleto{+}{5pt}}}\,,\\[5pt]
        \eta &= -\frac{1}{2} - \frac{\ell (\ell +1)}{\beta_{\scaleto{+}{5pt}}}\,.
    \end{aligned} \label{Heun_Parameters_NH}
\end{equation}
Using the relation \cref{R_u_relation}, together with the results of \cref{u_of_x_NH,Metric_Function_NH}, we obtain that:
\begin{equation}
	\begin{aligned}
        R(x) &= x^{(\beta -1)/2}\,
        (x-1)^{\gamma/2} \left[c_{1}\,\text{HeunC}\left(-\alpha,\,\gamma,\,\beta,\,-\delta,\,\delta+\eta;\,x\right) \right.  \\[5pt] 
        &+ \left. c_{2}\, \left(x-1\right)^{-\gamma}\,\text{HeunC}\left(-\alpha, \,-\gamma, \,\beta,\,-\delta, \,\delta+\eta; \,x\right) \right]\,, 
	\end{aligned} \label{R_WF_Complete}
\end{equation}
where $c_{1}$ and $c_{2}$ incorporated the remaining multiplicative constants.
	
Given that our solution is valid only near the event horizon (as $x \to 1^{+}$), we now investigate the behavior of $R(x)$ under the same limit. As detailed in the appendix \ref{app-Frobenius-Heun}, the confluent Heun function has a series expansion around $x=1$, with respect to the characteristic exponent $r_{1}=0$, that is valid for $1<x<2$, yielding:
\begin{equation}
	\text{HeunC}(-\alpha,\,\gamma,\, \beta,\,-\delta,\,\delta+\eta;\,x) = \sum_{n=0}^{+\infty} c_{n} \left(x-1\right)^{n}\,,
\end{equation}
where the coefficients $c_{n}$ satisfy the recurrence relations in \cref{Termo_Geral}, with $c_{0}=1$. In addition, an equivalent result can be shown for the case associated with $r_{2}=-\gamma$.

Thus, when $x \to 1^{+}$, we consider only the leading-order term of both series, approximating $\text{HeunC}(-\alpha,\,\pm\gamma,\,\beta,\,-\delta,\,\delta+\eta;\,x) \approx 1$. Applying this approximation to \cref{R_WF_Complete} yields the near-horizon solution:
\begin{equation}
        R_{\scaleto{NH}{4pt}}(x) = x^{(\beta-1)/2} \left[c_{1}\,\left(x-1\right)^{\gamma/2} +  c_{2}\, \left(x-1\right)^{-\gamma/2}\right]\,,    
        \label{radial-x-1}
\end{equation}
where the label \enquote{\textsc{nh}} emphasizes that this solution is accurate only as $x\to 1^{+}$, i.e., near the horizon.

The Heun parameters $\gamma$ and $\beta$ are essential to this solution, especially for the cases where $\alpha_{\scaleto{Q}{4pt}} = 1/2$ and $1$. While the quintessence parameter $N_{\scaleto{Q}{4pt}}$ has a macroscopically suppressed contribution to the event horizon radii (as shown in \cref{Event_Horizon_Formation}), its influence on the radial solutions is significant. This is because $N_{\scaleto{Q}{4pt}}$ is a key component of the $\beta_{\scaleto{+}{5pt}}$ parameter (defined in \cref{def_Beta_Plus}), which directly interferes in both $\gamma$ and $\beta$. These two parameters will play a crucial role in the analysis of \enquote{dark phases} presented in the next section.
	
Additionally, for future convenience, we rewrite \cref{radial-x-1} in the exponential form:
\begin{equation}
    \begin{aligned}
       R_{\scaleto{NH}{4pt}}(x) 
       &= c_{1}\,\exp \left[\frac{\beta}{2} (x-1) + \frac{\gamma}{2} \ln (x-1)\right]\\
       &+ c_{2}\,\exp \left[\frac{\beta}{2} (x-1) - \frac{\gamma}{2} \ln (x-1)\right]\,,
    \end{aligned} \label{R_WF_for_DPs}
\end{equation}
Observe that, in the result above, we used $\ln{x} \approx (x-1)$ as $x\to 1^{+}$. Notably, unless $|\beta| \gg |\gamma|$, we see that the dominant terms in \cref{R_WF_for_DPs} are the ones where
\begin{equation}
    R_{\scaleto{NH}{4pt}}(x) \propto (x-1)^{\pm\gamma/2}\,,
\end{equation}
which indicates that the real part of the wave function is proportional to $\cos \left[\frac{\gamma}{2} \ln (x-1)\right]$. The imaginary part differs by a $\pi/2$ shift.

The next section will explore the behavior of \cref{R_WF_for_DPs} and analyze the distinct features introduced into the radial solution for different values of the quintessence parameter $N_{\scaleto{Q}{4pt}}$.

\section{Dark Phases}\label{Section_DPs}

To investigate the quantum implications of dark energy (in the form of Kiselev's quintessential fluid) in scalar particles, we now explore the emergence of quintessence-induced dark phases in the radial wave function $R(x)$, determined by \cref{R_WF_for_DPs}. We continue to explore the scenarios of $\alpha_{\scaleto{Q}{4pt}} = 0,\,1/2,\,1$, along with physical constraints on $\overline{a}$, $r_{\scaleto{S}{4pt}}$, and $N_{\scaleto{Q}{4pt}}$. This analysis reveals how quintessence leaves an imprint on scalar particle dynamics, distinguishing the wave functions for different values of the quintessence parameter $N_{\scaleto{Q}{4pt}}$. Here, we only consider scenarios where $r_{\scaleto{S}{4pt}}\neq 0$; otherwise, there would be no event horizon formation.
    
From \cref{Heun_Parameters_NH,R_WF_for_DPs}, we see how the Heun parameters $\gamma$ and $\beta$ are fundamental in determining the behavior of the solution shown in \cref{R_WF_for_DPs}, valid in the vicinity of the event horizon. For this reason, we will have to determine them explicitly for each choice of $\alpha_{\scaleto{Q}{4pt}}$. To this aim, it is convenient to use results of \cref{Event-Horizon-Alpha_Q-0,EH-without-quintessence}, together with the general expression of $\beta_{\scaleto{+}{5pt}}$, given by \eqref{def_Beta_Plus}, to obtain the following results:
    \begin{table}[!ht]
	\begin{center}
		\begin{tabular}{ | c | c | c |}
			\hline
				$\alpha_{\scaleto{Q}{4pt}}$ & $r_{\scaleto{+}{5pt}}$ & $\beta_{\scaleto{+}{5pt}}$ \\
			\hline\hline
				$0$ & $r_{\scaleto{S}{4pt}}/(1+ \overline{a} + N_{\scaleto{Q}{4pt}})$ & $1 + \overline{a} + N_{\scaleto{Q}{4pt}}$ \\
				
				$1/2$ & $r_{\scaleto{S}{4pt}}/(1 + \overline{a})$ & $(1+\overline{a}) \left[1 + 2\,N_{\scaleto{Q}{4pt}} r_{\scaleto{S}{4pt}}/ (1+\overline{a})^{2}\right]$ \\
				
				$1$ & $r_{\scaleto{S}{4pt}}/(1 + \overline{a})$ & $(1+\overline{a}) \left[1 + 3\,N_{\scaleto{Q}{4pt}} r_{\scaleto{S}{4pt}}^{2}/ (1+\overline{a})^{3}\right]$\\
			\hline
		\end{tabular}
	\end{center}
    \caption{General results of $r_{\scaleto{+}{5pt}}$ and $\beta_{\scaleto{+}{5pt}}$ for scenarios with $\alpha_{\scaleto{Q}{4pt}} = 0,\,1/2,\,1$. The results in the second row ($\alpha_{\scaleto{Q}{4pt}} = 1/2$) are valid in the physical regime of $N_{\scaleto{Q}{4pt}}\,r_{\scaleto{S}{4pt}}/ (1+\overline{a})^{2} \ll 1$, while those of the third row are valid when $N_{\scaleto{Q}{4pt}} \,r_{\scaleto{S}{4pt}}^{2}/ (1+\overline{a})^{3} \ll 1$. Notably, these conditions are easily met when we consider our estimates for $N_{\scaleto{Q}{4pt}}$, given by \cref{N_Q_function_Alpha_Q}.}
    \label{table_r_plus_beta_plus}
    \end{table}
    
According to \cref{table_r_plus_beta_plus}, we can define the following cases, based on the values of $\overline{a}$, $r_{\scaleto{S}{4pt}}$, and $N_{\scaleto{Q}{4pt}}$: the standard cases (where all the parameters are non-zero), the cloudless cases (setting $\overline{a} = 0$), and the restricted case without quintessence (setting $N_{\scaleto{Q}{4pt}} = 0$). On the other hand, if we choose a configuration where $\overline{a} = N_{\scaleto{Q}{4pt}} = 0$, we fall into the original Schwarzschild case with $r_{\scaleto{+}{5pt}} = r_{\scaleto{s}{4pt}}$ and $\beta_{\scaleto{+}{5pt}} = 1$. Notably, our following analysis of dark phases is focused on standard cases, thereby encompassing all relevant contributions.

Having examined the general results for $r_{\scaleto{+}{5pt}}$ and $\beta_{\scaleto{+}{5pt}}$, we are able to determine the explicit form of the spin$-0$ particle's radial wave function near the event horizon for $\alpha_{\scaleto{Q}{4pt}} = 0,\,1/2,\,1$. By doing so, we will discuss the emergence of dark phases, which are similar to our previous results in black string spacetimes \cite{Deglmann2025}.

\subsection{Dark phases at the lower bound} \label{subDP0}

To determine the explicit form of the radial wave function $R(x)$ for $\alpha_{\scaleto{Q}{4pt}} = 0$, we first calculate the Heun parameters $\gamma$ and $\beta$, defined by \cref{Heun_Parameters_NH}. Using the results from \cref{table_r_plus_beta_plus}, we obtain that:
\begin{equation}
    \begin{aligned}
        \frac{\gamma}{2} 
        &= \pm i\, \frac{\epsilon\,r_{\scaleto{s}{4pt}}}{\left(1 + \overline{a} + N_{\scaleto{Q}{4pt}}\right)^{2}}\,,\\
        \frac{\beta}{2} 
        &= \pm i\,\left[\frac{\ell (\ell +1)}{1+\overline{a}+N_{\scaleto{Q}{4pt}}} - \frac{1}{4}\right]^{1/2}\,.
    \end{aligned}\label{Gamma_Beta_Alpha_Q_0}
\end{equation}
Hence, the radial wave function $R(x)$ can be written as
\begin{equation}
    \begin{aligned}
       R(x) 
       &= c_{\scaleto{+}{5pt}} \exp\left[i\,\delta_{+}^{\scaleto{(0)}{6pt}}(x)\right] + 
       c_{\scaleto{-}{4pt}} \exp\left[i\,\delta_{-}^{\scaleto{(0)}{6pt}}(x)\right]\,,
    \end{aligned} \label{RWF_DP_0}
\end{equation}
where $\delta_{\pm}^{\scaleto{(0)}{6pt}}(x)$ are given by
\begin{equation}
    \delta_{\pm}^{\scaleto{(0)}{6pt}}(x) = \pm \frac{\epsilon\,r_{\scaleto{s}{4pt}}}{\left(1 + \overline{a} + N_{\scaleto{Q}{4pt}}\right)^{2}} \ln(x-1) + \left[\frac{\ell (\ell +1)}{1+\overline{a}+N_{\scaleto{Q}{4pt}}} - \frac{1}{4}\right]^{1/2} (x-1)\,.\label{DP_0}
\end{equation}
The upper index \enquote{$\scaleto{(0)}{8pt}$} associates the functions $\delta_{\pm}^{\scaleto{(0)}{6pt}}(x)$ to their value of $\alpha_{\scaleto{Q}{4pt}} = 0$. Moreover, $\delta_{\pm}^{\scaleto{(0)}{6pt}}(x)$ encodes the effect of quintessence in the radial wave function $R(x)$ so that we name it the \emph{dark phase}. Since $\delta_{\pm}^{\scaleto{(0)}{6pt}}(x)$ are not directly proportional to $N_{\scaleto{Q}{4pt}}$, we could use the term \emph{implicit dark phase} to this case. This happens because, from \cref{Contribution_Numeric_Estimate}, we expect $N_{\scaleto{Q}{4pt}} \approx 8.02$ in the lower bound of $\alpha_{\scaleto{Q}{4pt}}$. For such a value of $N_{\scaleto{Q}{4pt}}$, we cannot expand the expression to put the quintessence parameter in evidence.

\begin{figure}[ht!]
        \begin{center}
            \includegraphics[width=0.65\textwidth]{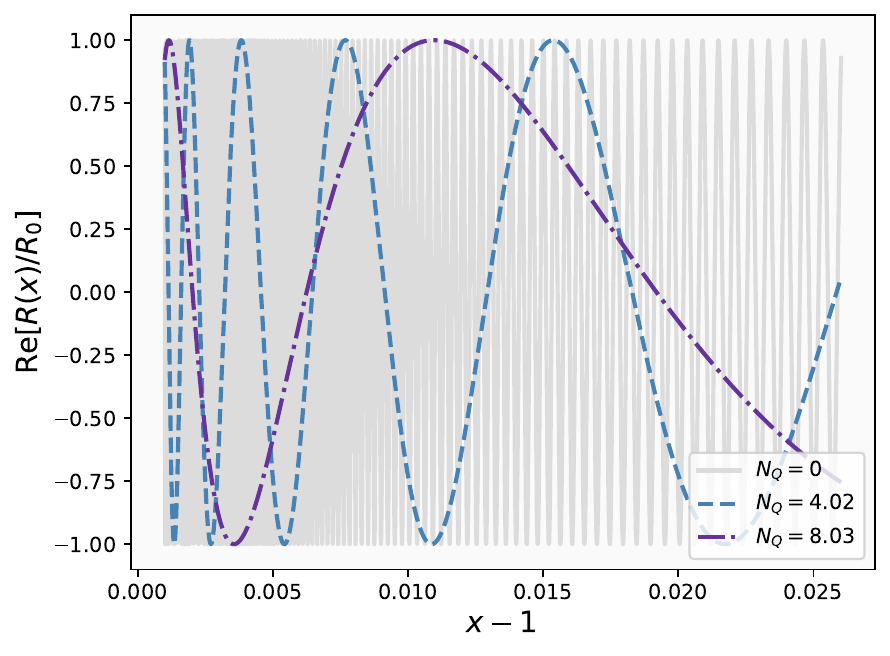}
            \caption{Real part of the non-normalized radial wave function $R(x)$, given by \cref{RWF_DP_0,DP_0}, as a function of the dimensionless radial coordinate $x = r/r_{\scaleto{+}{5pt}}$ for $\alpha_{\scaleto{Q}{4pt}} = 0$. The parameters are fixed at $r_{\scaleto{s}{4pt}} = 10^{14}$ m, $\overline{a} = 10^{-6}$, $\epsilon = 2\,\overline{m}_{\phi}$, $l=3$ and $\overline{m}_{\phi} = m_{\phi} c/\hbar$, with $m_{\phi} = 4 \times 10^{-55}$ kg. Note that besides the compression of these oscillations as $x\to 1^{+}$, due to $\ln (x-1)$, the wavelength increases with increasing $N_{\scaleto{Q}{4pt}}$.}
            \label{Fig_DP_0}
        \end{center}
    \end{figure}

\begin{figure}[ht!]
    \centering
    \minipage{0.49\textwidth}
        \includegraphics[width=\linewidth]{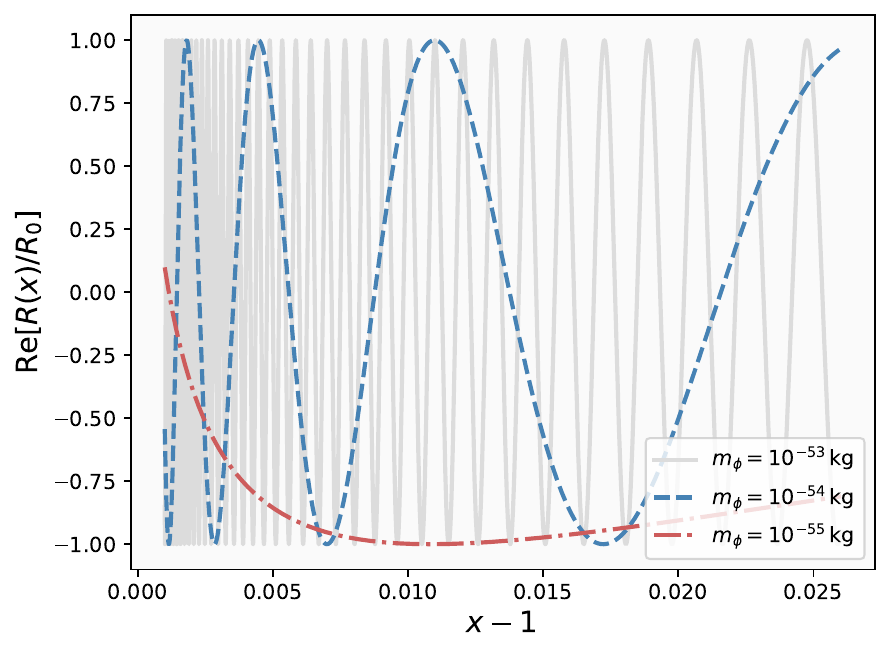}
    \endminipage\hfill
    \minipage{0.49\textwidth}%
        \includegraphics[width=\linewidth]{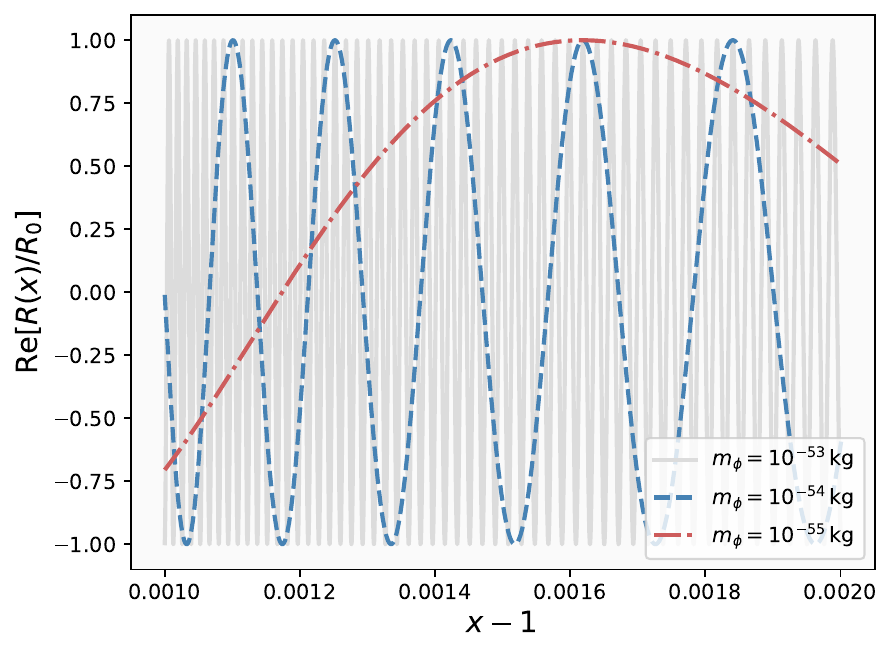}
    \endminipage
    \caption{Real part of the non-normalized radial wave function given by \cref{RWF_DP_0,DP_0} for different scalar particle masses. The fixed parameters are set as $r_{\scaleto{s}{4pt}} = 10^{14}$ m, $\overline{a} = 10^{-6}$, $\epsilon = 2\,\overline{m}_{\phi}$, $\overline{m}_{\phi} = m_{\phi} c/\hbar$ and $l=3$. On the left, we consider the estimated quintessence parameter for $\alpha_{\scaleto{Q}{4pt}}=0$, $N_{\scaleto{Q}{4pt}}=8.03$, while the right panel is depicted with $N_{\scaleto{Q}{4pt}}=2.41$.}
    \label{DP_0_masses}
\end{figure}

\begin{figure}[ht!]
  \centering
  \begin{minipage}{0.48\linewidth}
        \centering
        \subcaptionbox{}
        {\includegraphics[width=\linewidth, trim={0 {15pt} 0 {15pt}},clip]
        {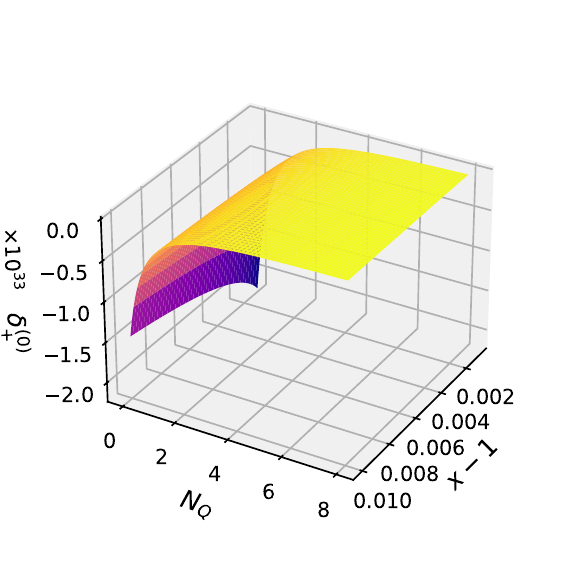}}
    \end{minipage}
    \quad
    \begin{minipage}{0.48\linewidth}
        \centering
        \subcaptionbox{}    
        {\includegraphics[width=\linewidth, trim={0 {15pt} 0 {15pt}},clip]
        {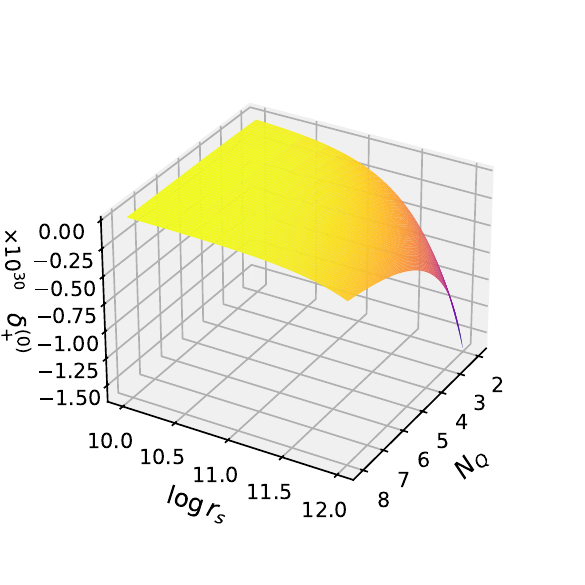}}
    \end{minipage}
    
    \begin{minipage}{0.48\linewidth}
        \centering
        \subcaptionbox{}
        {\includegraphics[width=\linewidth, trim={0 {15pt} 0 {15pt}},clip]{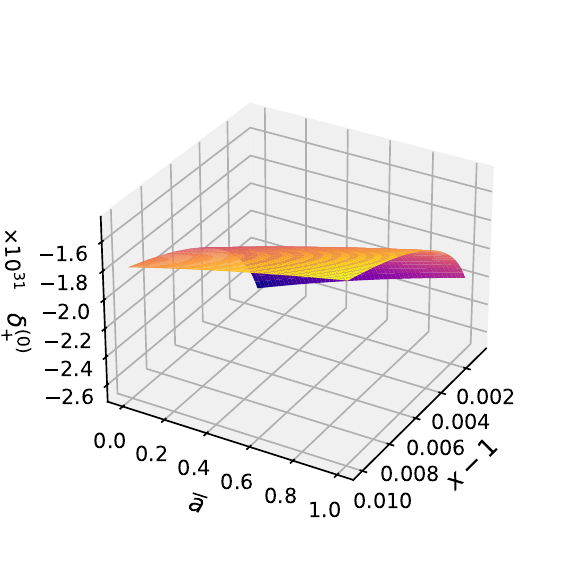}}
    \end{minipage}  
    \quad
    \begin{minipage}{0.48\linewidth}
        \centering
        \subcaptionbox{}
        {\includegraphics[width=\linewidth, trim={0 {15pt} 0 {15pt}},clip]{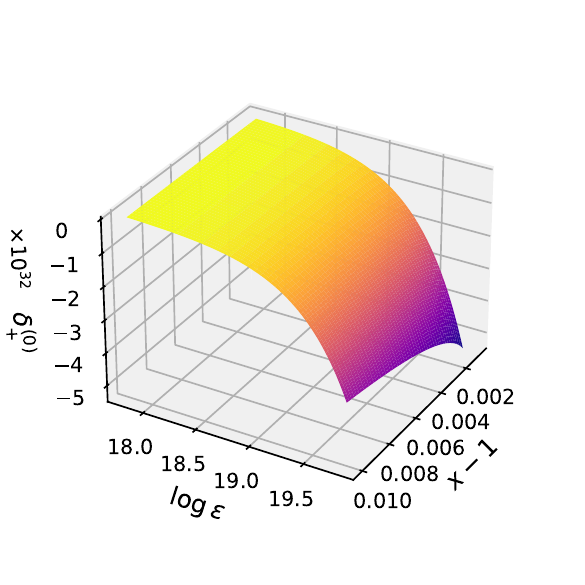}}
    \end{minipage}
 \caption{Dark phase at the lower bound, represented by \cref{DP_0}, as a function of spacetime and particle parameters. For all fixed values, we set $r_{\scaleto{S}{4pt}}=10^{14}$ m, $\overline{a} = 10^{-3}$, $x = 1.01$ m, $\ell = 3$ and $\epsilon = 5 \, \overline{m}_{\phi}$. Furthermore, we considered $\overline{m}_{\phi}$ to be the normalized Higgs mass, where $m_{\phi} = 2.24 \times 10^{-25}$ kg (see \cref{def_Epsilon_M_bar}), and $N_{\scaleto{Q}{4pt}}$ is defined according to \cref{N_Q_Estimate} with $r_{\text{obs}} = 4.4 \times 10^{26}$ m.}
\label{3D_DP_0}
\end{figure}

For completeness, we illustrate the effect of $N_{\scaleto{Q}{4pt}}$ on $R(x)$ in \cref{Fig_DP_0}. As $x \rightarrow{1^{+}}$, the $\text{ln}(x-1)$ term substantially compresses the real part of the non-normalized radial wave function, and this effect becomes even more relevant for small values of $N_{\scaleto{Q}{4pt}}$. Additionally, \cref{Fig_DP_0} presents the induced phase for a scalar particle with mass $m_{\phi} = 4 \times 10^{-55}$ kg ($\approx 2.49 \times 10^{-22}$ $e \text{V}/c^{2}$), which is a mass scale typically considered for ultralight axion-like particles. For higher mass particles, the left panel of \cref{DP_0_masses} shows that the oscillations are much faster, and this effect is even more pronounced for smaller values of $N_{\scaleto{Q}{4pt}}$, as presented in the right panel for $N _{\scaleto{Q}{4pt}}=2.41$. The impact of the quintessence parameter on the phase is further illustrated in \cref{3D_DP_0}, where we consider the spin$-0$ boson mass to be in the Higgs scale. As depicted in the top-left panel, a decrease in $N_{\scaleto{Q}{4pt}}$ causes $\delta_{+}^{\scaleto{(0)}{6pt}}(x)$ to become more pronounced, manifesting as more rapid oscillations in the radial function.

\Cref{3D_DP_0}  additionally presents the dark phase at the top-right panel as a function of the Schwarzschild radius and the quintessence parameter. Notably, while increasing the order of $r_{\scaleto{S}{4pt}}$, the phase decreases very rapidly as $N_{\scaleto{Q}{4pt}}$ diminishes. This decay is mitigated as $N_{\scaleto{Q}{4pt}}$ approaches its expected value. For fixed $N_{\scaleto{Q}{4pt}}$, on the other hand, changes in the cloud parameter yield minor variations of the phase, as shown in the bottom left panel of \cref{3D_DP_0}. This effect is expected when examining the metric function, given by \cref{Metric_Function}. At the lower bound, the quintessence term loses its radial dependence and becomes comparable to the cloud parameter. As a result, $N_{\scaleto{Q}{4pt}}$ becomes dominant over small values of $\overline{a}$ in \cref{DP_0}, thereby suppressing the cloud contribution. 
\begin{figure}
    \centering
    \includegraphics[width=0.6\linewidth, trim={0 {15pt} 0 {50pt}},clip]{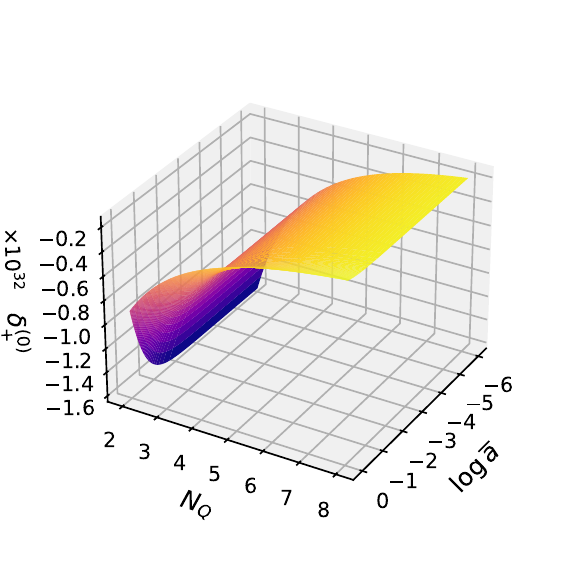}
    \caption{Behavior of the dark phase at the lower bound, given by \cref{DP_0}, as a function of the quintessence and cloud parameters. We set $r_{s}=10^{14}$ m, $x = 1.01$ m, $\ell = 3$ and $\epsilon = 5 \, \overline{m}_{\phi}$. The normalized mass is considered according to \cref{def_Epsilon_M_bar} with  $m_{\phi} = 2.24 \times 10^{-25}$ kg.}
    \label{3D_DP_0_NQ_a_bar}
\end{figure}

\Cref{3D_DP_0_NQ_a_bar} provides further insight into how the phase behaves as the quintessence and cloud parameters vary. As previously stated, $N_{\scaleto{Q}{4pt}}$ significantly dominates at smaller values of $\overline{a}$, amplifying the phase when the quintessence parameter decreases and diminishing it when $N_{\scaleto{Q}{4pt}}$ increases. In contrast, higher values of the cloud parameter mitigate the influence of small $N_{\scaleto{Q}{4pt}}$, resulting in less intense phases. This interplay of the cloud and quintessence parameters is analogous to what is observed in horizon formation. As depicted in the left panel of \cref{rHorizon_0_Meio},  $\overline{a}$ has significant impact only at its higher values and for low $N_{\scaleto{Q}{4pt}}$.

Finally, we highlight an interesting feature of \cref{DP_0}. While panels (a), (b) and (c) in \cref{3D_DP_0} incorporate the choice $\epsilon = 5 \, \overline{m}_{\phi}$ for the spin$-0$ particle, the bottom right panel depicts the dark phase without imposing a specific mass-dependent energy relation. For higher orders of $\epsilon$, the phase becomes progressively intense, implying a decrease in the wavelength as oscillations become faster.

\subsection{Dark phases at the middle of the interval}

\begin{figure}[ht!]
    \begin{center}
        \includegraphics[width=0.65\textwidth]{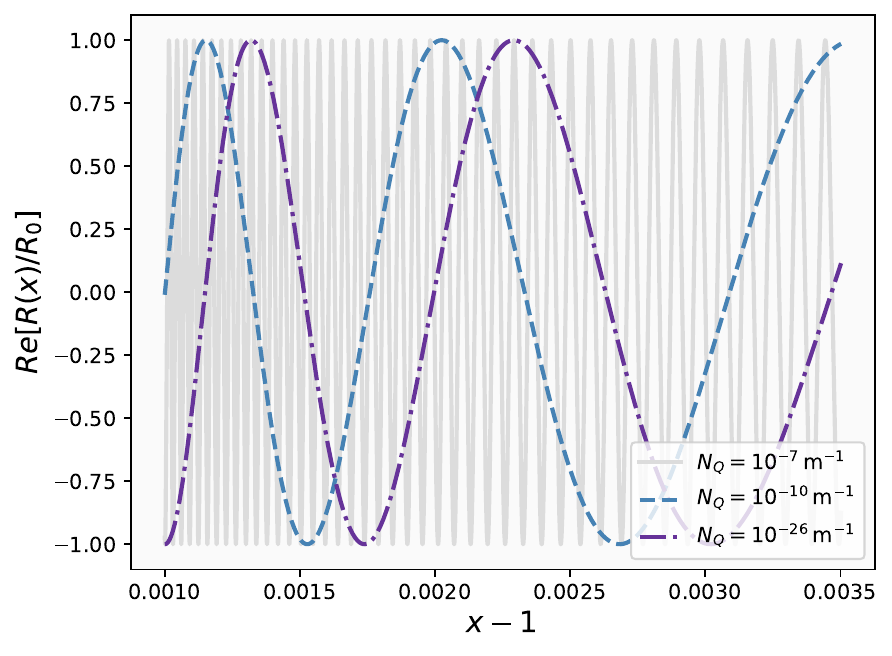}
        \caption{Real part of the non-normalized radial wave function $R(x)$, given by \cref{RWF_Meio_Intervalo,Phi_Meio_Intervalo,DP_Meio_Intervalo}, as a function of  $x=r/r_{\scaleto{+}{5pt}}$, for $\alpha_{\scaleto{Q}{4pt}} = 1/2$. The parameters are fixed at $r_{\scaleto{s}{4pt}} = 10^{8}$ m, $\overline{a} = 10^{-6}$, $\epsilon = 4\,\overline{m}_{\phi}$, $l=3$ and $\overline{m}_{\phi} = m_{\phi} c/\hbar$, with $m_{\phi}=10^{-50}$ kg. Note that the quintessence-induced phase shift increases with $x$.}
        \label{DP_Meio_NH}
    \end{center}
\end{figure}

Our next task is to investigate the occurrence of dark phases when $\alpha_{\scaleto{Q}{4pt}} = 1/2$ (corresponding to $\omega_{\scaleto{Q}{4pt}} = -2/3$). To do so, we use the results from the second row of \cref{table_r_plus_beta_plus} to determine the Heun parameters $\gamma$ and $\beta$, given by \cref{Heun_Parameters_NH}, explicitly. This procedure yields:
\begin{equation}
    \begin{aligned}
        \frac{\gamma}{2} 
        &= \pm i\, \frac{\epsilon\,r_{\scaleto{s}{4pt}}}{\left(1 + \overline{a}\right)^{2}}\left[1 - \frac{2\,N_{\scaleto{Q}{4pt}} r_{\scaleto{S}{4pt}}}{(1+\overline{a})^{2}}\right]\,,\\[5pt]
        \frac{\beta}{2} 
        &= \pm i\,\left[\frac{\ell(\ell+1)}{1+\overline{a}}-\frac{1}{4}\right]^{1/2}
        \mp i\,\frac{\ell(\ell+1)}{(1+\overline{a})}\left[\frac{\ell(\ell+1)}{1+\overline{a}}-\frac{1}{4}\right]^{-1/2} \frac{N_{\scaleto{Q}{4pt}}\,r_{\scaleto{s}{4pt}}}{(1+\overline{a})^{2}}\,,
    \end{aligned}\label{Gamma_Beta_Alpha_Q_Meio}
\end{equation}
where we present a first-order approximation of $\beta/2$ since $N_{\scaleto{Q}{4pt}}\,r_{\scaleto{s}{4pt}}/(1+\overline{a})^{2}$ is of the order of $10^{-11}$ or smaller. Then, substituting the results of \cref{Gamma_Beta_Alpha_Q_Meio} into the radial wave function of \cref{R_WF_for_DPs}, we show that $R(x)$ can be written as
\begin{equation}
    R(x) = \Phi_{+}(x)\,\exp\left[-i\,\delta_{+}^{\scaleto{(1/2)}{6pt}}(x)\right] + \Phi_{-}(x)\,\exp\left[-i\,\delta_{-}^{\scaleto{(1/2)}{6pt}}(x)\right]\,,\label{RWF_Meio_Intervalo}
\end{equation}
where
\begin{equation}
    \Phi_{\pm}(x) = c_{\pm} \exp\left[\pm i\, \frac{\epsilon\,r_{\scaleto{s}{4pt}}}{\left(1 + \overline{a}\right)^{2}} \ln (x-1) + i\,\sqrt{\frac{\ell(\ell+1)}{1+\overline{a}}-\frac{1}{4}} \, (x-1)\right]\,\label{Phi_Meio_Intervalo}
\end{equation}
is independent of $N_{\scaleto{Q}{4pt}}$. The dark phases $\delta_{\pm}^{\scaleto{(1/2)}{6pt}}(x)$ are determined by:
\begin{equation}
    \delta_{\pm}^{\scaleto{(1/2)}{6pt}}(x) = \left[\pm \frac{2\,\epsilon\,r_{\scaleto{s}{4pt}}}{\left(1 + \overline{a}\right)^{2}}\, \ln (x-1) + \frac{\ell(\ell+1)}{(1+\overline{a})}\sqrt{\frac{\ell(\ell+1)}{1+\overline{a}}-\frac{1}{4}}^{-1} (x-1)\right]   
    \frac{N_{\scaleto{Q}{4pt}}\,r_{\scaleto{S}{4pt}}}{(1+\overline{a})^{2}}\,. \label{DP_Meio_Intervalo}
\end{equation}
Therefore, the scalar particle's radial wave function is indeed modified in the presence of our dark energy candidate, and this modification is an explicit phase that increases with $x$. We show a numerical example of this case in \cref{DP_Meio_NH}, where the quintessence-induced phase shift is shown for our expected value of $N_{\scaleto{Q}{4pt}}$, according to \cref{N_Q_function_Alpha_Q}, and also for larger values of $N_{\scaleto{Q}{4pt}}$ (for comparison). For smaller quintessence parameters and $N_{\scaleto{Q}{4pt}}=0$ m$^{-1}$ (spacetime without quintessence), the small distances between the curves presents numerical challenges, limiting their comparison with $N_{\scaleto{Q}{4pt}} = 10^{-26}$ m$^{-1}$.

\begin{figure}[ht!]
    \centering
    \minipage{0.49\textwidth}
        \includegraphics[width=\linewidth]{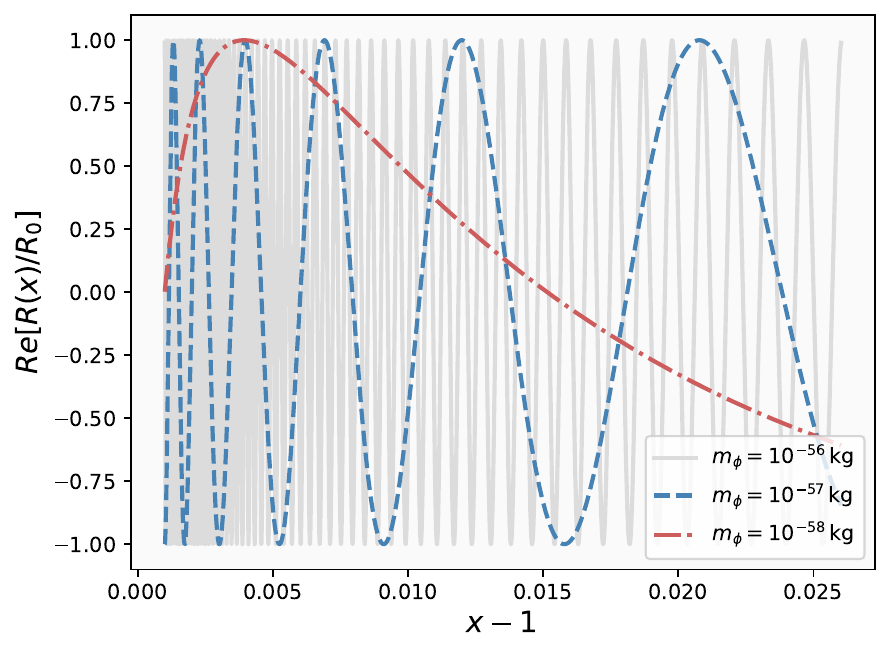}
    \endminipage\hfill
    \minipage{0.49\textwidth}%
        \includegraphics[width=\linewidth]{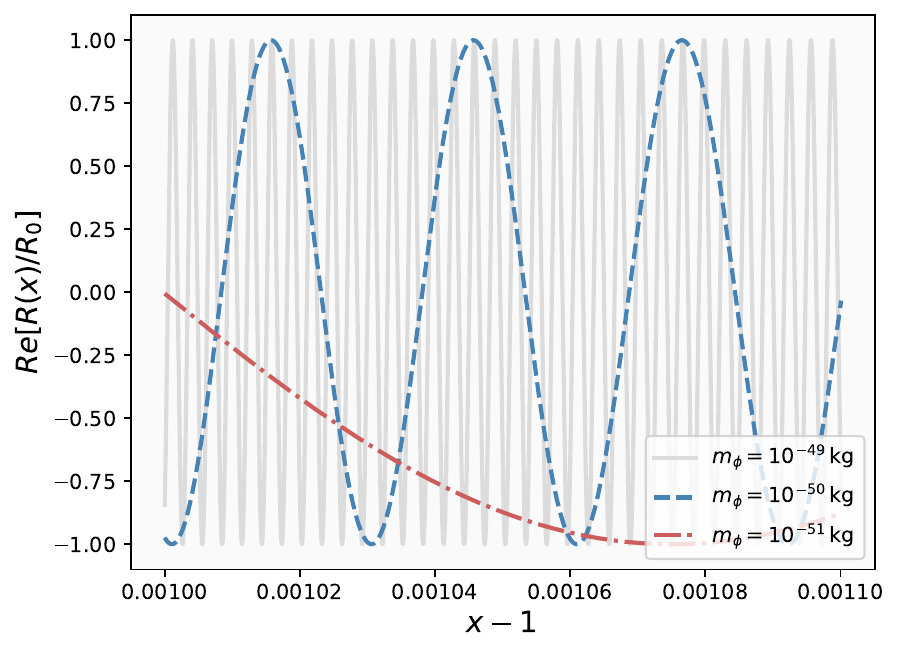}
    \endminipage
    \caption{Real part of the non-normalized radial wave function $R(x)$, given by \cref{RWF_Meio_Intervalo,Phi_Meio_Intervalo,DP_Meio_Intervalo}, for  different particle masses. The parameters are set to $r_{\scaleto{s}{4pt}} = 10^{8}$ m, $\overline{a} = 10^{-6}$, $\epsilon = 4\,\overline{m}_{\phi}$ and $l=3$. The left panel depicts the estimated quintessence parameter $N_{\scaleto{Q}{4pt}}=10^{-26}$ $\text{m}^{-1}$, while the right panel accounts for $N_{\scaleto{Q}{4pt}}=10^{-7}$ $\text{m}^{-1}$.}
    \label{DP_Meio_masses}
\end{figure}

Moreover, \cref{DP_Meio_masses} illustrates that lighter scalar particles exhibit a wider wavelength, while heavier particles oscillate more rapidly. This outcome is a direct consequence of the choice $\epsilon = 4\,\overline{m}_{\phi}$, since higher $\epsilon$ amplifies the compression caused by the $\ln(x-1)$ term in \cref{DP_Meio_Intervalo}. In addition, while the left panel depicts the phase within a scenario with the estimated quintessence parameter ($N_{\scaleto{Q}{4pt}}=10^{-26}$ $\text{m}^{-1}$), the right panel presents a supersaturated regime ($N_{\scaleto{Q}{4pt}}=10^{-7}$ $\text{m}^{-1}$). As shown in \cref{DP_Meio_NH}, larger quintessence parameters lead to faster oscillations, and this effect is even more pronounced for higher mass particles. This is the opposite of the behavior observed in $\alpha_{\scaleto{Q}{4pt}}=0$, where a decrease in the value of $N_{\scaleto{Q}{4pt}}$ enhance the phase (see \cref{DP_0_masses}).

\begin{figure}[ht!]
  \centering
  \begin{minipage}{0.48\linewidth}
        \centering
        \subcaptionbox{}
        {\includegraphics[width=\linewidth, trim={0 {15pt} 0 {15pt}},clip]{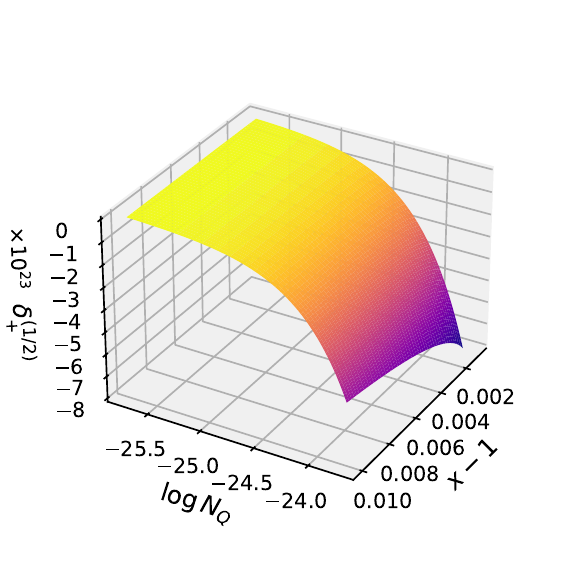}}   
    \end{minipage}
    \quad
    \begin{minipage}{0.48\linewidth}
        \centering
        \subcaptionbox{}    
        {\includegraphics[width=\linewidth, trim={0 {15pt} 0 {15pt}},clip]{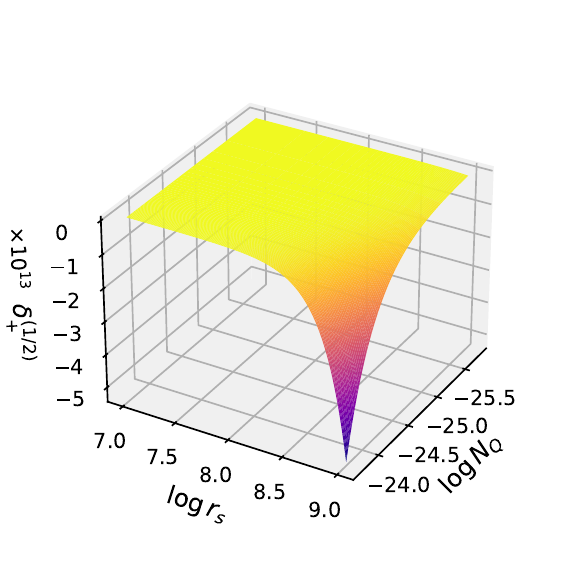}}
    \end{minipage}
    
    \begin{minipage}{0.48\linewidth}
        \centering
        \subcaptionbox{}
        {\includegraphics[width=\linewidth, trim={0 {15pt} 0 {15pt}},clip]{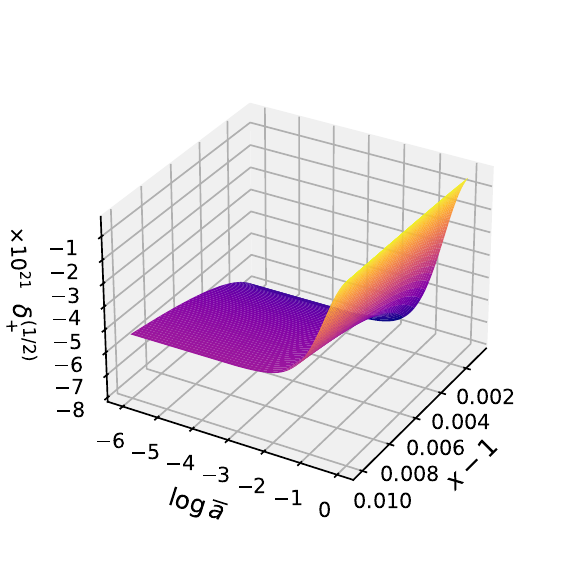}}
    \end{minipage}  
    \quad
    \begin{minipage}{0.48\linewidth}
        \centering
        \subcaptionbox{}
        {\includegraphics[width=\linewidth, trim={0 {15pt} 0 {15pt}},clip]{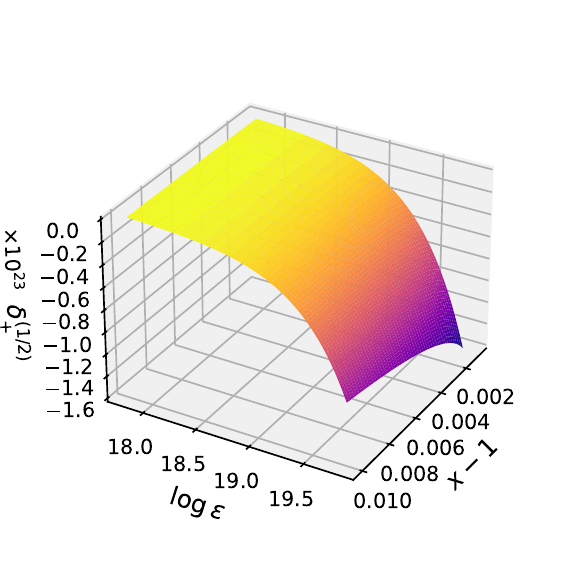}}
    \end{minipage}
  \caption{Dark phase at the middle of the interval, given by \cref{DP_Meio_Intervalo}, as a function of spacetime and particle parameters. For all fixed values, we set $r_{\scaleto{S}{4pt}}=10^{14}$ m, $\overline{a} = 10^{-3}$, $x = 1.01$ m, $\ell = 3$ and $\epsilon = 5 \, \overline{m}_{\phi}$. Furthermore, we considered $\overline{m}_{\phi}$ to be the normalized Higgs mass, where $m_{\phi} = 2.24 \times 10^{-25}$ kg (see \cref{def_Epsilon_M_bar}), and $N_{\scaleto{Q}{4pt}}$ is defined according to \cref{N_Q_Estimate} with $r_{\text{obs}} = 4.4 \times 10^{26}$ m.}
  \label{3D_DP_Meio}
\end{figure}

To further analyze the behavior of \cref{DP_Meio_Intervalo}, we illustrate in \cref{3D_DP_Meio} the dark phase as a function of spacetime and particle parameters considering the Higgs mass. As shown in the top-left panel, increasing the quintessence parameter enhances the phase, leading to more rapid oscillations. The Schwarzschild radius also plays a crucial role. Panel (a) employs $r_{\scaleto{S}{4pt}}=10^{14}$ m, whereas the top-right panel displays the phase evolution while for varying $r_{\scaleto{S}{4pt}}$ and $N_{\scaleto{Q}{4pt}}$. As predicted by \cref{DP_Meio_Intervalo}, a larger $r_{s}$ yields more intense phases, which can be observed comparing panels (a) and (b).

\begin{figure}
    \centering
    \includegraphics[width=0.6\linewidth, trim={0 {15pt} 0 {50pt}},clip]{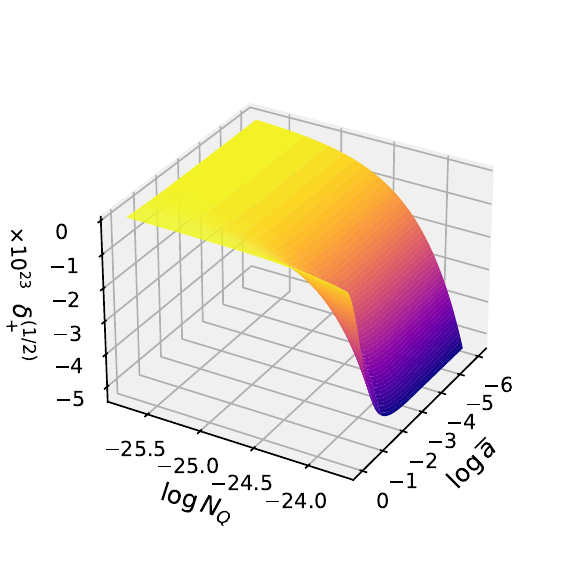}
    \caption{Dark phase at the middle of the interval, represented by \cref{DP_Meio_Intervalo}, as a function of $N_{\scaleto{Q}{4pt}}$ and $\overline{a}$. We set the Schwarzschild radius to $r_{\scaleto{S}{4pt}}=10^{14}$ m, $x = 1.01$ m, $\ell = 3$ and $\epsilon = 5 \, \overline{m}_{\phi}$, while $\overline{m}_{\phi}$ is the normalized Higgs mass, given by \cref{def_Epsilon_M_bar} with $m_{\phi} = 2.24 \times 10^{-25}$ kg.}
    \label{3D_DP_Meio_NQ_a_bar}
\end{figure}

Within our estimated range for $\overline{a}$, we further expect, from \cref{DP_Meio_Intervalo}, that increasing the cloud parameter will reduce the compressive effect of the $\text{ln}(x-1)$ term on the radial wave function. Concurrently, higher values of the cloud parameter reduce the influence of $N_{\scaleto{Q}{4pt}}$. The bottom left panel clearly illustrates this combined effect. Near the horizon, the phase remains reasonably steady for $\overline{a} < 10^{-2}$. As $\overline{a}\to 1$, a steep increase in $\delta_{+}^{\scaleto{(1/2)}{4pt}}$ occurs, and more intense clouds will exhibit less intense dark phases, thereby producing slower oscillations. This effect is also illustrated in \cref{3D_DP_Meio_NQ_a_bar}. As $\overline{a}$ decreases, $N_{\scaleto{Q}{4pt}}$ regulates the phase intensity, with higher quintessence parameter values leading to more pronounced dark phases. Higher clouds, on the other hand, counteract the $N_{\scaleto{Q}{4pt}}$ contribution, moderating the phase and the oscillations. Additionally, we observe that for $\overline{a} < 10^{-2}$, this relation is the opposite of that shown in the right panel of \cref{rHorizon_0_Meio}. While the horizon's position heavily depends on the cloud parameter, the radial behavior of scalar particles in its vicinity is more sensitive to changes in the quintessence parameter.

Moreover, the bottom-right panel displays the dark phase as a function of the spin-0 boson's energy. As $\epsilon$ increases, the phase becomes notably intense, implying that the real part of the radial function is more compressed at higher energies. This behavior is consistent with that observed in \cref{3D_DP_0} for the lower bound scenario. Nevertheless, \cref{3D_DP_Meio} reveals that the phase shown in \cref{DP_Meio_Intervalo} is considerably smaller in magnitude when compared to that in \cref{DP_0}. In fact, the dark phase produced at $\alpha_{\scaleto{Q}{4pt}}=1/2$ exhibits an overall moderate phase compared to that at $\alpha_{\scaleto{Q}{4pt}}=0$.

For the lower bound, no expansions were made in the quintessence parameter\footnote{Note that $N_{\scaleto{Q}{4pt}}\approx8.02$.}. However, at the middle of the interval, the expected value of $N_{\scaleto{Q}{4pt}}$ is much smaller ($N_{\scaleto{Q}{4pt}}=10^{-26}$ $\text{m}^{-1}$), enabling the regime where $N_{\scaleto{Q}{4pt}}\,r_{\scaleto{s}{4pt}}/(1+\overline{a})^{2} \ll 1$. Within this regime, \cref{DP_Meio_Intervalo} expresses a direct suppression of the $\text{ln}(x-1)$ contribution near the horizon, consequently yielding a less intense phase.

Finally, we emphasize that higher-energy spin$-0$ particles can overcome this suppression, leading to more intense phases. This is evident in the scenario where $\epsilon$ varies, as depicted in the bottom-right panel of \cref{3D_DP_Meio}. For a specific choice of mass-dependent energy relation, such as $\epsilon=5\, \overline{m}_{\phi}$, this also holds, so that heavier particles yield more intense phases.

\subsection{Dark phases in the upper bound}

Finally, we consider the case of $\alpha_{\scaleto{Q}{4pt}} = 1$ (or, equivalently, $\omega_{\scaleto{Q}{4pt}}=-1$), where quintessence makes its subtlest contribution to the metric function $f(r)$. Nevertheless, this is the most interesting regime, given the current observational constraints on the dark energy equation of state \cite{PlanckCosmology2014,DESI62024,DESI2025}.

We begin by determining the explicit Heun parameters $\gamma$ and $\beta$ of \cref{Heun_Parameters_NH}. Subsequently, we calculate the final expression for the radial wave function and its resulting dark phase. Using the results from the third row of \cref{table_r_plus_beta_plus}, in conjunction with \cref{Heun_Parameters_NH}, we show that:
\begin{equation}
    \begin{aligned}
        \frac{\gamma}{2} 
        &= \pm i\, \frac{\epsilon\,r_{\scaleto{s}{4pt}}}{\left(1 + \overline{a}\right)^{2}}\left[1 - \frac{3\,N_{\scaleto{Q}{4pt}} r_{\scaleto{S}{4pt}}^{2}}{(1+\overline{a})^{3}}\right]\,,\\[5pt]
        \frac{\beta}{2} 
        &= \pm i\,\left[\frac{\ell(\ell+1)}{1+\overline{a}}-\frac{1}{4}\right]^{1/2}
        \mp i\,\frac{3}{2}\,\frac{\ell(\ell+1)}{(1+\overline{a})}\left[\frac{\ell(\ell+1)}{1+\overline{a}}-\frac{1}{4}\right]^{-1/2} \frac{N_{\scaleto{Q}{4pt}}\,r_{\scaleto{s}{4pt}}^{2}}{(1+\overline{a})^{3}}\,.
    \end{aligned}\label{Gamma_Beta_Alpha_Q_1}
\end{equation}
Substituting these expressions for $\gamma$ and $\beta$ into the radial wave function given by \cref{R_WF_for_DPs}, yields:
\begin{equation}
    R(x) = \Phi_{+}(x)\,\exp\left[-i\,\delta_{+}^{\scaleto{(1)}{6pt}}(x)\right] + \Phi_{-}(x)\,\exp\left[-i\,\delta_{-}^{\scaleto{(1)}{6pt}}(x)\right]\,,\label{RWF_Upper_Bound}
\end{equation}
where $\Phi_{\pm}(x)$ has the exact same expression as \cref{Phi_Meio_Intervalo}, while the dark phases $\delta_{\pm}^{\scaleto{(1)}{6pt}}$ are determined by:
\begin{equation}
    \delta^{\scaleto{(1)}{6pt}}_{\pm}(x) = \left[\pm \frac{3\,\epsilon\,r_{\scaleto{s}{4pt}}}{\left(1 + \overline{a}\right)^{2}}\, \ln (x-1) + \frac{3}{2}\, \frac{\ell(\ell+1)}{(1+\overline{a})}\sqrt{\frac{\ell(\ell+1)}{1+\overline{a}}-\frac{1}{4}}^{-1} (x-1)\right]   
    \frac{N_{\scaleto{Q}{4pt}}\,r_{\scaleto{S}{4pt}}^{2}}{(1+\overline{a})^{3}}\,. \label{DP_Upper_Bound}
\end{equation}

\begin{figure}[ht!]
    \begin{center}
        \includegraphics[width=0.65\textwidth]{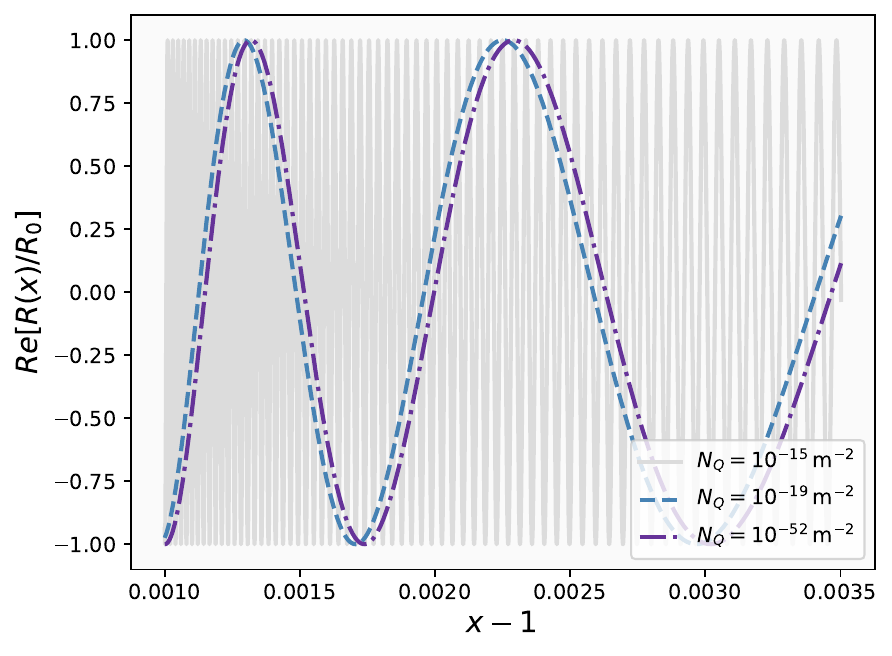}
        \caption{Real part of the non-normalized radial wave function $R(x)$, given by \cref{RWF_Upper_Bound,Phi_Meio_Intervalo,DP_Upper_Bound}, as a function of $(x-1)$, where $x = r/r_{\scaleto{+}{5pt}}$, for $\alpha_{\scaleto{Q}{4pt}} = 1$. The parameters are fixed at $r_{\scaleto{s}{4pt}} = 10^{8}$ m, $\overline{a} = 10^{-6}$, $\epsilon = 4\,\overline{m}_{\phi}$, and $l=3$. Additionally, $\overline{m}_{\phi}$ is the normalized mass given by \cref{def_Epsilon_M_bar} with $m_{\phi} = 10^{-50}$ kg.}
        \label{DP_1_NH}
    \end{center}
\end{figure}

As seen from \cref{DP_Upper_Bound}, these dark phases increase with the distance $x$. However, since $N_{\scaleto{Q}{4pt}}\,r_{\scaleto{S}{4pt}}^{2}/(1+\overline{a})^{3}$ is of the order $10^{-22}$ or smaller, it is challenging to \enquote{measure} the induced phase difference between $N_{\scaleto{Q}{4pt}} = 0$ m$^{-2}$ (when quintessence is absent) and $N_{\scaleto{Q}{4pt}} = 10^{-52}$ m$^{-2}$ (when it is present). For this reason, our numerical example in \cref{DP_1_NH} considers three values for $N_{\scaleto{Q}{4pt}}$: the first two representing a Universe supersaturated with dark energy, and the third case being the physically admissible one. However, note that for $N_{\scaleto{Q}{4pt}}=10^{-19}$ $\text{m}^{-2}$ the radial function approaches the curve with the estimated quintessence parameter ($N_{\scaleto{Q}{4pt}}=10^{-52}$ $\text{m}^{-2}$). This is expected since the term $N_{\scaleto{Q}{4pt}}\,r_{\scaleto{S}{4pt}}^{2}/(1+\overline{a})^{3}$ in \cref{DP_Upper_Bound} becomes progressively smaller as $N_{\scaleto{Q}{4pt}}$ decreases.

\begin{figure}[ht!]
    \centering
    \minipage{0.49\textwidth}
        \includegraphics[width=\linewidth]{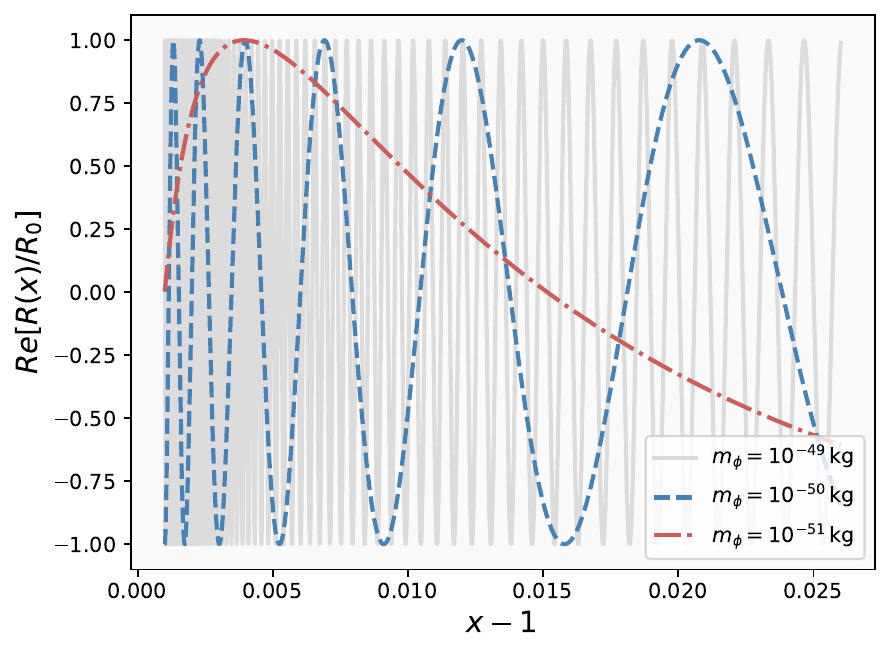}
    \endminipage\hfill
    \minipage{0.49\textwidth}%
        \includegraphics[width=\linewidth]{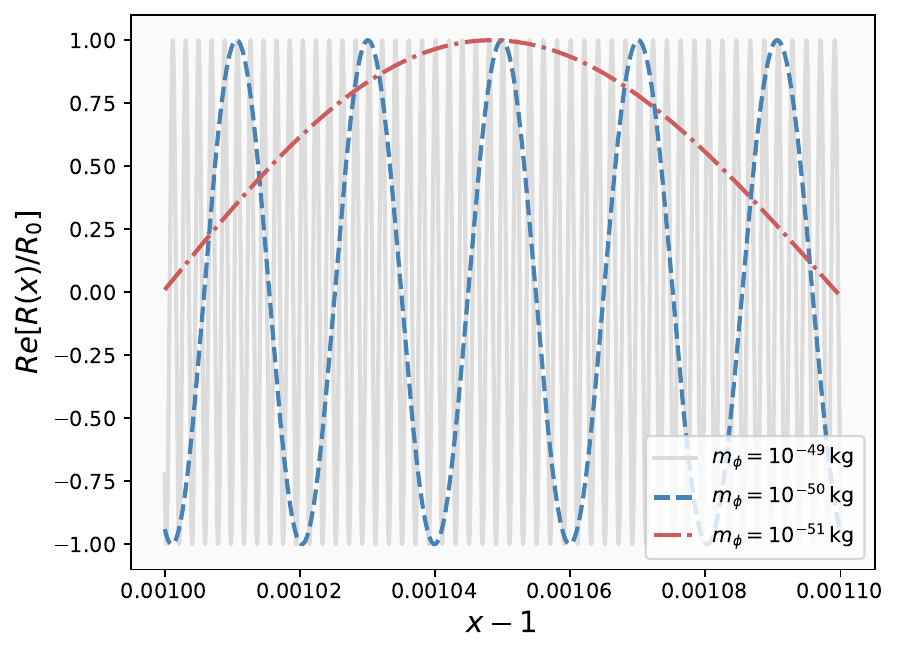}
    \endminipage
    \caption{Real part of the non-normalized radial wave function $R(x)$, given by \cref{RWF_Upper_Bound,Phi_Meio_Intervalo,DP_Upper_Bound}, considering different masses for the spin$-0$ particle. We set the parameters at $r_{\scaleto{s}{4pt}} = 10^{8}$ m, $\overline{a} = 10^{-6}$, $\epsilon = 4\,\overline{m}_{\phi}$ and $l=3$. On the left panel, the quintessence parameter is set to $N_{\scaleto{Q}{4pt}}=10^{-52}$ $\text{m}^{-2}$, while the right panel illustrates the real wave functions with $N_{\scaleto{Q}{4pt}}=10^{-15}$ $\text{m}^{-2}$.}
    \label{DP_1_masses}
\end{figure}

\Cref{DP_1_NH} further presents the real part of the wave function for a scalar particle with mass $m_{\phi} = 10^{-50}$ kg ($\sim 10^{-18}$ $e\text{V}/c^{2}$). For higher order masses, the left panel of \cref{DP_1_masses} shows that the wavelength is considerably small, and this effect is enhanced when the quintessence parameter increases (as depicted in the right panel). Additionally, since the estimated quintessence parameters for $\alpha_{\scaleto{Q}{4pt}}=1/2, \, 1$ differentiate themselves in a extremely small regime, we point to the fact that the real radial wave functions depicted in the left panel of \cref{DP_1_masses} are identical to those in the left panel \cref{DP_Meio_masses}.

\begin{figure}[ht!]
  \centering
    \begin{minipage}{0.48\linewidth}
        \centering
        \subcaptionbox{}
        {\includegraphics[width=\linewidth, trim={0 {15pt} 0 {15pt}},clip]
        {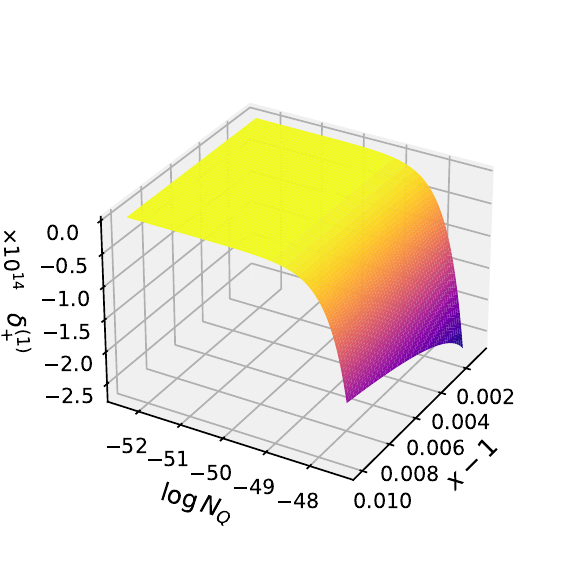}}   
    \end{minipage}
    \quad
    \begin{minipage}{0.48\linewidth}
        \centering
        \subcaptionbox{}{\includegraphics[width=\linewidth, trim={0 {15pt} 0 {15pt}},clip]
        {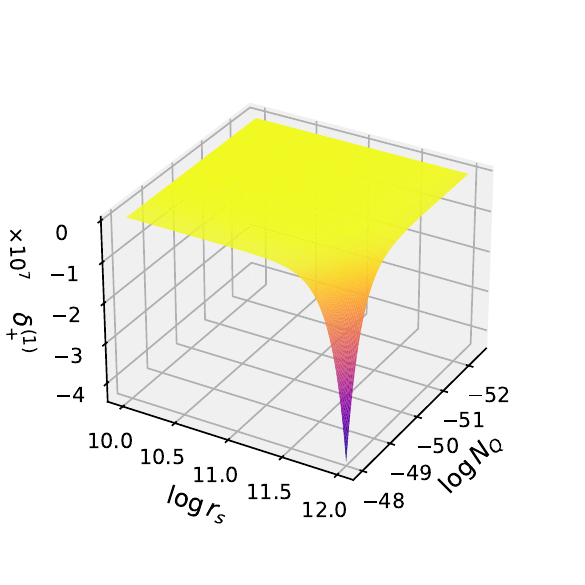}}
    \end{minipage}

    \begin{minipage}{0.48\linewidth}
        \centering
        \subcaptionbox{}    {\includegraphics[width=\linewidth, trim={0 {15pt} 0 {15pt}},clip]{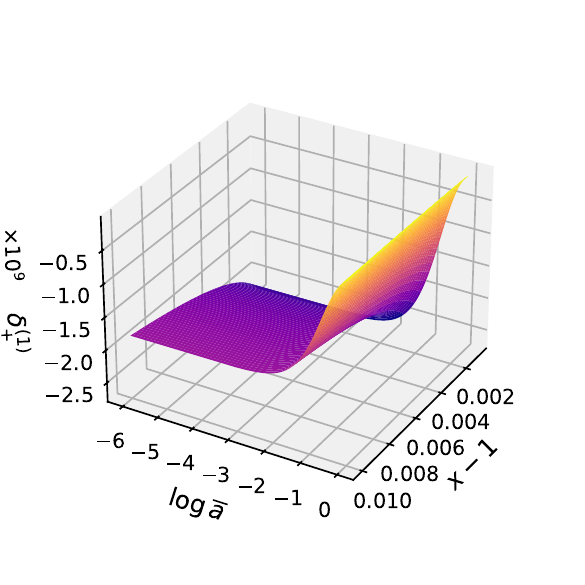}}   
    \end{minipage}
    \quad
    \begin{minipage}{0.48\linewidth}
        \centering
        \subcaptionbox{}
        {\includegraphics[width=\linewidth, trim={0 {15pt} 0 {15pt}},clip]{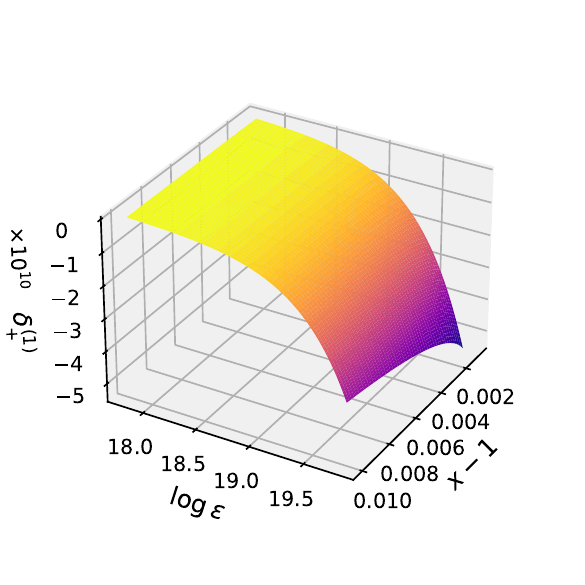}}
  \end{minipage}  
  \caption{Dark phase at the upper bound, given by \cref{DP_Upper_Bound}, as a function of spacetime and particle parameters. For all fixed values, we set $r_{\scaleto{S}{4pt}}=10^{14}$ m, $\overline{a} = 10^{-3}$, $x = 1.01$ m, $\ell = 3$ and $\epsilon = 5 \, \overline{m}_{\phi}$. Furthermore, we considered $\overline{m}_{\phi}$ to be the normalized Higgs mass, where $m_{\phi} = 2.24 \times 10^{-25}$ kg (see \cref{def_Epsilon_M_bar}), and $N_{\scaleto{Q}{4pt}}$ is defined according to \cref{N_Q_Estimate}.}
  \label{3D_DP_1}
\end{figure} 
Moreover, the top-left panel of \cref{3D_DP_1} depicts the dark phase as a function of the quintessence parameter $N_{\scaleto{Q}{4pt}}$ considering the Higgs mass. Consistent with our previous analysis for $\alpha_{\scaleto{Q}{4pt}}=1/2$, the phase becomes progressively intense as $N_{\scaleto{Q}{4pt}}$ increases. In addition, the top-right panel illustrates the same behavior upon simultaneous increases in $N_{\scaleto{Q}{4pt}}$ and $r_{s}$. A key difference, compared to \cref{3D_DP_Meio}, is evident in the order of magnitude of $\delta_{+}^{\scaleto{(1)}{6pt}}$.

\begin{figure}
    \centering
    \includegraphics[width=0.6\linewidth, trim={0 {15pt} 0 {50pt}},clip]{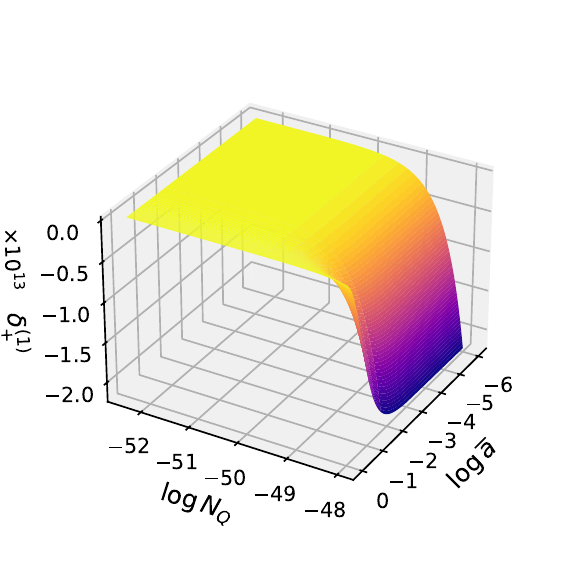}
    \caption{Behavior of the dark phase at the upper bound, represented by \cref{DP_Upper_Bound}, as a function of $N_{\scaleto{Q}{4pt}}$ and $\overline{a}$. The fixed parameters are $r_{\scaleto{S}{4pt}}=10^{14}$ m, $x = 1.01$ m, $\ell = 3$ and $\epsilon = 5 \, \overline{m}_{\phi}$. The normalized mass, defined by \cref{def_Epsilon_M_bar}, was set with the Higgs mass $m_{\phi} = 2.24 \times 10^{-25}$ kg.}
    \label{3D_DP_1_NQ_a_bar}
\end{figure}

Concerning the cloud parameter, we expect the phase to be less intense for larger values of $\overline{a}$. This is because larger cloud parameters not only mitigate the effect of the logarithmic term in \cref{DP_Upper_Bound}, but also minimize the global contribution from $N_{\scaleto{Q}{4pt}}\,r_{s}/(1+\overline{a})^{3}$. \Cref{3D_DP_1_NQ_a_bar} illustrates this effect, highlighting the dominance of the $N_{\scaleto{Q}{4pt}}$ parameter when $\overline{a}$ is reduced. Similar to the dark phase when $\alpha_{\scaleto{Q}{4pt}}=1/2$ (given by \cref{DP_Meio_Intervalo} and represented in \cref{3D_DP_Meio_NQ_a_bar}), for small values of the cloud parameter, the order of the quintessence parameter dictates the phase intensity. This leads to more rapid oscillations as  $N_{\scaleto{Q}{4pt}}$ increases and slower oscillations as $N_{\scaleto{Q}{4pt}}$ decreases. Conversely, larger clouds reduce this effect, moderating the phase as $\overline{a} \rightarrow 1$.

Nevertheless, in $\delta^{\scaleto{(1)}{6pt}}_{+}$ we find this effect to be considerably more subtle than in $\delta^{\scaleto{(1/2)}{6pt}}_{+}$, with the quintessence parameter dominating across nearly all estimated cloud values. We also observe that, unlike the horizon position, the dark phase for $\alpha_{\scaleto{Q}{4pt}}=1$ is more affected by changes in the quintessence parameter than in the cloud parameter. This is clear when comparing \cref{3D_DP_1_NQ_a_bar} and \cref{rHorizon_1}.

Furthermore, \cref{3D_DP_1} illustrates how the dark phase behaves with increasing energy $\epsilon$ (bottom-right panel). For the case of quintessence at the upper bound ($\alpha_{\scaleto{Q}{4pt}}=1$), higher-energy spin-0 particles enhance the phase's magnitude, consistent with the behavior at $\alpha_{\scaleto{Q}{4pt}}=0$ and also $\alpha_{\scaleto{Q}{4pt}}=1/2$.

At last, the phases derived for $\alpha_{\scaleto{Q}{4pt}}=1$ are considerably smaller in magnitude than those analyzed previously. This is explicitly shown in \cref{DP_Upper_Bound}, where the term $N_{\scaleto{Q}{4pt}}\,r_{s}^{2}/(1+\overline{a})^{3} \ll 1$ suppresses the $\text{ln}(x-1)$ term near the horizon, consequently leading to a less intense phase compared to \cref{DP_0} and \cref{DP_Meio_Intervalo}.

\section{Conclusions} \label{Section 7_conclusions}

The emergence of dark phases within the dynamics of scalar particles highlights an interesting aspect of dark energy's influence in the quantum framework. As our previous work demonstrated \cite{Deglmann2025}, these dark phases are intrinsically dependent on the quintessence parameter $N_{\scaleto{Q}{4pt}}$, enabling us to search for observables and investigate the physical consequences of different dark energy state parameters.

Building upon these previous analyses, this work investigated the influence of the quintessence fluid on spin$-0$ particles. In particular, we considered scalar particles in the vicinity of a Schwarzschild-Kiselev-Letelier black hole horizon, providing detailed solutions to the Klein-Gordon equation for three specific values of the dark energy state parameter ($\alpha_{\scaleto{Q}{4pt}}=0, \, 1/2, \, 1$). To achieve this, we first introduced the exact solution to Einstein's field equations for our specific spacetime and analyzed its effects on event horizon formation. As described in \cref{Event_Horizon_Formation}, for $\alpha_{\scaleto{Q}{4pt}}=1/2$ and $1$, the dark energy parameter becomes significantly small, suppressing the quintessential term in the metric function for distances $r\ll r_{\scaleto{obs.}{4pt}}$. This effect results in the dominance of the Schwarzschild radius in horizon formation. While this holds true, we also note that the contribution of $N_{\scaleto{Q}{4pt}}$ at the quantum level is significant. As demonstrated in subsequent sections, $N_{\scaleto{Q}{4pt}}$ emerges as a significant factor for the radial solutions due to its relationship with other relevant parameters.

In \cref{KG-equation}, we provided a detailed analytical solution to the Klein-Gordon equation near the event horizon for the lower bound of the state parameter. By employing a first-order approximation to the metric function, we also presented solutions to the Klein-Gordon equation for the lower, upper, and intermediate bounds of $\alpha_{\scaleto{Q}{4pt}}$. In each scenario, the corresponding solution took the form of a Confluent Heun Function, which we discussed in \cref{app-confluent-Heun}, along with properties of the Confluent Heun Equation.

Finally, in \cref{Section_DPs}, we analyzed the results from \cref{KG-equation}. The solutions yielded a unique dark phase, which we investigated for its dependence on the theory’s parameters; the results are shown in Figures $9$--$20$. As observed, for some values of these parameters, the dark phases may exhibit more significant effects. Furthermore, we highlight that the phases are strictly tied to the quintessence parameter. Its existence provides a way to distinguish the presence of quintessence from its absence, as well as from different values of $N_{\scaleto{Q}{4pt}}$, thus allowing it to be interpreted as a relative phase difference.

Another crucial finding from this section was that these dark phases, in spherically symmetric spacetime, are much smaller and subtler than our previous results in cylindrical spacetimes. It appears that, beyond the effect of our dark energy candidate, the geometry of the spacetime itself influences whether these dark phases are more or less difficult to be \enquote{measured}. This novel behavior needs to be investigated further in a subsequent work.

We further emphasize that the emergence of dark phases within the scalar particle solutions affects well-known results of quantum field theory, including Hawking radiation \cite{Parikh_2000} and the Unruh effect, which are related to particle production. For instance, the dependence of the Hawking temperature on these phases can be determined and investigated; an analysis that we defer to future work.

The Hanbury Brown and Twiss (HBT) effect \cite{HBT1956, baym1998, Hama1988}, which is observed in a wide range of experiments from stellar astronomy to heavy-ion collisions, provides an additional opportunity for observing dark phases. Since some of these experiments are based on the interferometry of quantum particles produced with random phases, the HBT effect is well-suited for this task. The primary challenge, however, is to identify systems with parameter configurations where the effect is significant enough to be measured. This possibility requires further investigation.

\section*{Acknowledgments}

The authors are grateful for the valuable contributions of the reviewers, who raised significant issues and contributed positively to an improved version of this work. B.V.S. was supported by the Conselho Nacional de Desenvolvimento Científico e Tecnológico (CNPq), grant number 141549/2023-8. M.L.D. would like to thank the Coordenação de Aperfeiçoamento de Pessoal de Nível Superior – Brasil (CAPES) – Finance Code 001. C.C.B.Jr. thanks CNPq, for financial support.

\appendix
\section{Confluent Heun functions}\label{app-confluent-Heun}

To properly characterize the physical systems within our study, this appendix presents the confluent case of the Heun equation and its associated local solutions, which are related to the Klein-Gordon radial equations detailed in \Cref{KG-equation}.

We start by considering the canonical form of the confluent Heun equation (CHE), which is given by \cite{Olver2010}:
\begin{equation}
    \frac{d^{2} y}{d z^{2}} + \left(\alpha + \frac{\beta + 1}{z} + \frac{\gamma + 1}{z - 1}\right)\frac{d y}{d z} + \left(\frac{\mu}{z} + \frac{\nu}{z - 1}\right)y = 0\,,\label{HeunConfluentCanonical}
\end{equation}
which has regular singularities at $z=0$ and $z=1$, and an irregular singularity at $z\to\infty$. The parameters $\mu$ and $\nu$ are determined in terms of the Heun parameters, $\alpha$, $\beta$, $\gamma$, $\delta$, and $\eta$, by:
\begin{align}
    \mu &= \frac{1}{2}\left(\alpha - \beta - \gamma + \alpha\beta - \beta\gamma\right) - \eta\,,\label{Mu_Condition}\\
    \nu &= \frac{1}{2}\left(\alpha + \beta + \gamma + \alpha\gamma + \beta\gamma\right) + \delta + \eta\,.\label{Nu_Condition}
\end{align}
Notably, Mathieu functions, spheroidal wave functions, and Coulomb spheroidal functions are special cases of solutions of the confluent Heun equation \cite{Olver2010}. The solutions to the CHE belong to three classes: the local solutions (obtained with the Fuchs-Frobenius method), the Heun functions (where the local solutions are extended to encompass two regular singularities), and the Heun polynomials (with an extended domain).

As we detail in the next section, a \emph{local} solution can be determined about a regular singular point $z_{0} = 0,\,1$ using the Fuchs-Frobenius method \cite{Olver2010}. The associated solution will be valid within the radius of convergence, which for the CHE is $|z-z_{0}|<1$. These results will be useful when dealing with radial Klein-Gordon solutions.

\subsection{Fuchs-Frobenius solution to the confluent Heun equation}\label{app-Frobenius-Heun}

We now explore the Fuchs-Frobenius solution of the confluent Heun equation \cref{HeunConfluentCanonical} about the regular singularity lying at $z=1$. We call this solution \emph{local} because the regular singular points of \cref{HeunConfluentCanonical} imply a finite radius of convergence of exactly one unit. Thus, our \emph{local} results will be valid for $|z-1|<1$. Local solutions can be extended by analytic continuation, but we do not discuss this issue here. For more details, please see \cite{Fiziev2010}.

Firstly, let us multiply \cref{HeunConfluentCanonical} by a factor of $z(z-1)$, so that
\begin{equation}
    \begin{aligned}
        0 &= z\left(z-1\right) y''(z)\\ 
        &+ \left[\alpha z\left(z - 1\right) + \left(\beta + 1\right)\left(z - 1\right) + \left(\gamma+1\right) z\right] y'(z)\\ 
        &+ \left[\mu\left(z-1\right) + \nu z\right] y(z)\,.
    \end{aligned} \label{Confluent_Eq_Aberta}
\end{equation}
For convenience, we define the auxiliary variable $\chi = z - 1$, implying that $d/d\chi = d/d z$. In terms of $\chi$, \cref{Confluent_Eq_Aberta} becomes
\begin{equation}
    \begin{aligned}
        0 &= \chi^{2} \,y'' + \Theta_{c}\,\chi \,y' + \nu \,y\\
        &+ \chi \,y'' + \left(\gamma + 1\right) y'\\
        &+ \alpha \,\chi^{2} \,y' + \left(\mu + \nu\right) \chi \,y\,,
    \end{aligned} \label{Eq_para_chi}
\end{equation}
where we had defined $\Theta_{c}$ as
\begin{equation}
\Theta_{c} = \alpha + \beta + \gamma + 2\,.\label{Omega_c}
\end{equation}
Additionally, the prime notation denotes a derivative with respect to $\chi$, i.e., $d/d \chi$.
	
Moving forward, we substitute the general Frobenius' ansatz\footnote{The index $r$ should not be confused with a coordinate.}
\begin{equation}
    y_{a}(\chi) = \sum_{m=0}^{+\infty} c_{m}\,\chi^{m+r}\,,\label{series-chi}
\end{equation}
into \cref{Eq_para_chi} to show that
\begin{equation}
    \begin{aligned}
        0 &= \sum_{m=0}^{+\infty} c_{m} \left[\left(m+r\right)\left(m+r+\Theta_{c}-1\right) + \nu\right] \chi^{m+r}\\
        &+ \sum_{m=0}^{+\infty} c_{m} \left(m+r\right)\left(m+r+\gamma\right) \chi^{m+r-1}\\
        &+\sum_{m=0}^{+\infty} c_{m} \left[\alpha\left(m+r\right) + \left(\mu + \nu\right)\right] \chi^{m+r+1}\,.
    \end{aligned}\label{Confluent_z_1_case}
\end{equation}
In the next step, we rename the summation index $m$, of \cref{Confluent_z_1_case}, according to: $k = m$ in the first sum; $k = m-1$ in the second one; and $k = m+1$ for the last term. This yields:
\begin{equation}
    \begin{aligned}
        0 &= \sum_{k=0}^{+\infty} c_{k} \left[\left(k+r\right)\left(k+r+\Theta_{c}-1\right) + \nu\right] \chi^{k+r}\\
        &+ \sum_{k=-1}^{+\infty} c_{k+1} \left(k+r+1\right)\left(k+r+\gamma+1\right) \chi^{k+r}\\
        &+\sum_{k=1}^{+\infty} c_{k-1} \left[\alpha\left(k+r-1\right) + \left(\mu + \nu\right)\right] \chi^{k+r}\,.
    \end{aligned}\label{Confluent_Expanded_Around_1}
\end{equation}
By equating the coefficients of each power of $z$ to zero in the above equation, we obtain that
\begin{align}
    &c_{0} \,r \left(r + \gamma\right) = 0\,,\label{Equação_Indicial}\\
    &c_{1} = \frac{r\left(r + \alpha + \beta + \gamma + 1\right) + \nu}{\left(r+1\right)\left(r+\gamma+1\right)} \,c_{0}\,,
    \label{c_1_Confluente_1}
\end{align}
while the general recurrence term is
\begin{equation}
M_{0} \,c_{m} + M_{1} \,c_{m+1} + M_{2} \,c_{m+2} = 0\,, \quad m\geq 0\,,\label{Termo_Geral}
\end{equation}
with
\begin{equation}
    \begin{aligned}
        M_{0} &= \alpha\left(m+r\right) + \left(\mu+\nu\right)\,,\\
        M_{1} &= \left(m+r+1\right)\left(m+r+\alpha+\beta+\gamma+2\right) + \nu\,,\\
        M_{2} &= \left(m+r+2\right)\left(m+r+\gamma+2\right)\,.
    \end{aligned}\label{Ms}
\end{equation}
We observe that the first equation, namely \cref{Equação_Indicial}, is an indicial (or characteristic) equation, whose roots are $r=0$ and $r=-\gamma$. Then, provided that $\gamma$ is not zero or an integer, \cref{HeunConfluentCanonical} has two independent solutions 
\begin{equation}
    y_{a}(z) =\sum_{m=0}^{\infty}c_{m}\left(z-1\right)^{m+r}\,,\label{CHE_Fuchs_Frobenius_Solution}
\end{equation}
for $r=0$ and $r=-\gamma$ and $a=1,\,2$. We emphasize that the index $a$ in $y_{a}(z)$ should have been added to $r$ and $c_{m}$ in \cref{CHE_Fuchs_Frobenius_Solution}, to distinguish the two possibilities. However, we chose to suppress these indexes to avoid excessive notation.

In addition, \cref{c_1_Confluente_1} determines the coefficient $c_{1}$ for both cases (of $r=0$ and $r=-\gamma$), while the three-term recursion relation, described by \cref{Termo_Geral,Ms}, gives the remaining coefficients (again for $r=0$ and $r=-\gamma$).

Since all the coefficients $c_{m}$ depend on $c_{0} = 1$, it is a standard practice to set $c_{0}=1$. In this case, we can explicitly show the coefficient $c_{1}$ of the local solution \eqref{CHE_Fuchs_Frobenius_Solution} about $z=1$, which is given by
\begin{equation}
c_{1} = \frac{r\left(r + \alpha + \beta + \gamma + 1\right)}{\left(r+1\right)\left(r+\gamma+1\right)} + \frac{\left(\alpha + \beta\right)\left(1+\gamma\right)+ \gamma + 2 \left(\delta+\eta\right)}{2 \left(r+1\right)\left(r+\gamma+1\right)}\,.
\end{equation}
The subsequent coefficients follow from \cref{Termo_Geral,Ms}.

\subsubsection{Imposing the break off of the series solution}
	
According to the solution given by \cref{CHE_Fuchs_Frobenius_Solution,Equação_Indicial,c_1_Confluente_1,Termo_Geral,Ms}, one can demand the break off of the series \eqref{CHE_Fuchs_Frobenius_Solution} into a polynomial requiring that $M_{0} = 0$ and $c_{m+1}\left(\alpha,\,\beta,\,\gamma,\,\delta,\,\eta\right) = 0$ for some $m = N$, with $N\in \mathbb{Z^{+}}$.
	
The first requirement, i.e. $M_{0} = 0$ for some $m=N\in\mathbb{Z}^{+}$, gives us the condition
\begin{equation}
\alpha \left(N+1+\frac{\beta+\gamma}{2}\right) + \delta = 0\,.\label{Delta_Condition}
\end{equation} 
The above condition is known as the \enquote{$\delta$--condition} and it is useful when determining spectral constraints.

\subsection{Liouville normal form}\label{app-liouville-Heun}

In addition to our previous description, it is often convenient to express the CHE in the Liouville normal form \cite{Titchmarsh1962}. For this purpose, we write \cref{HeunConfluentCanonical} as
    \begin{equation}
	y''(z) + p(z)\, y'(z) + q(z)\, y(z) = 0, \label{General_2nd_Order_ODE}
    \end{equation}
where $y(z)$ is the solution to \cref{HeunConfluentCanonical}, while $p(z)$ and $q(z)$ are identified as
    \begin{align}
        p(z) &= \alpha + \frac{\beta + 1}{z} + \frac{\gamma + 1}{z - 1}\,,\label{p(z)}\\
        q(z) &= \frac{\mu}{z} + \frac{\nu}{z - 1}\,.\label{q(z)}
    \end{align}
We further introduce the functions $I(z)$ and $u(z)$. They are related by
    \begin{equation}
        y(z) = I(z)\,u(z)\,,\label{Ansatz_Normal_Form}
    \end{equation}
from which \cref{General_2nd_Order_ODE} can be expressed as
    \begin{align}
        0 &= I(z)\, u''(z) + \left[2 I'(z) + p(z) I(z)\right]u'(z)\notag\\[5pt]
        &+ \left[I''(z) + p(z) I'(z) + q(z) I(z)\right] u(z)\,. \label{Auxiliary_I}
    \end{align}
By setting the coefficient of $u'(z)$ to zero, we find that
    \begin{equation}
        I'(z) = -\frac{1}{2}\, p(z)\, I(z)\,.\label{Eq_I_Deriv}
    \end{equation}
Therefore, as long as $I(z) \neq 0$, \cref{Auxiliary_I} reads
    \begin{equation}
        u''(z) + \left[-\frac{1}{2} p'(z) - \frac{1}{4} p(z)^{2} + q(z)\right] u(z) = 0\,,\label{General_Equation_Liouville_Normal_Form}
    \end{equation}
which is the general normal form of a second-order linear ODE as \cref{General_2nd_Order_ODE}. Moreover, plugging \cref{p(z)} and \cref{q(z)} into \cref{General_Equation_Liouville_Normal_Form}, we see that
    \begin{equation}
        u''(z) + \left[\frac{A}{z} + \frac{B}{(z - 1)} + \frac{C}{z^{2}} + \frac{D}{(z - 1)^{2}} + E\right] u(z) = 0\,,\label{Eq_u_Normal_Form}
    \end{equation}
where the coefficients were set using \cref{Mu_Condition} and \cref{Nu_Condition}, and are defined as  
    \begin{equation}
        \begin{aligned}
            A &= \tfrac{1}{2} - \eta\,,\\
            B &= -\tfrac{1}{2} + \delta + \eta\,,\\
            C &= \left(1 - \beta^{2}\right)/4\,,\\
		D &= \left(1 - \gamma^{2}\right)/4\,,\\
		E &= -\alpha^{2}/4\,.
	\end{aligned} \label{Parameters_HeunC_Normal}
    \end{equation}
    
Finally, we recall that \cref{Ansatz_Normal_Form} provides a solution to \cref{Eq_u_Normal_Form} in terms of $I(z)$ and the standard Heun function. After integrating $I(z)$, we obtain the following solution (centered at $z=1$):
    \begin{equation}
	\begin{aligned}
            u(z) &= e^{\alpha z/2}\,z^{(1+\beta)/2}\,\left(z - 1\right)^{(1 + \gamma)/2} \left[c_{1}\,\text{HeunC} \left(\alpha,\,\gamma,\,\beta,\,-\delta,\,\delta+\eta;\,z\right) \right.\\
            & \left.+ c_{2}\,(z-1)^{-\gamma}\, \text{HeunC} \left(\alpha,\,-\gamma,\,\beta,\,-\delta,\,\delta + \eta;\,z\right)\right] \,, \label{Sol_Geral_u}
	\end{aligned}
    \end{equation}
where $c_{1}$ and $c_{2}$ are arbitrary constants. More details on Heun equations can be obtained from \cite{Olver2010}.

\bibliography{Artigo_3}
\bibliographystyle{unsrt} 

\end{document}